\documentclass[12pt]{article}

\usepackage{slashed}
\usepackage{amssymb,amsmath}
\usepackage{graphicx}

\DeclareMathOperator{\sech}{sech}
\DeclareMathOperator{\sgn}{sgn}

\begin{document}

\begin{titlepage}

\begin{flushright}
arXiv:2309.13221
\end{flushright}
\vskip 2.5cm

\begin{center}
{\Large \bf Creation of Bound Half-Fermion Pairs by Solitons}
\end{center}

\vspace{1ex}

\begin{center}
{\large Sapan Karki and Brett Altschul\footnote{{\tt altschul@mailbox.sc.edu}}}

\vspace{5mm}
{\sl Department of Physics and Astronomy} \\
{\sl University of South Carolina} \\
{\sl Columbia, SC 29208} \\
\end{center}

\vspace{2.5ex}

\medskip

\centerline {\bf Abstract}

\bigskip

In the presence of topologically nontrivial bosonic field configurations, the fermion number operator may
take on fractional eigenvalues, because of the existence of zero-energy fermion modes. The simplest examples
of this occur in $1+1$ dimensions, with zero modes attached to kink-type solitons. In the presence of a
kink-antikink pair, the two associated zero modes bifurcate into positive and negative energy levels with
energies $\pm ge^{-g\Delta}$, in terms of the Yukawa coupling $g\ll 1$ and the distance
$\Delta$ between the kink and antikink centers. When the kink and antikink are moving, it seems that
there could
be Landau-Zener-like transitions between these two fermionic modes, which would be interpretable as the
creation or annihilation of fermion-antifermion pairs; however, with only two solitons in relative motion,
this does not occur. If a third solitary wave is introduced farther away to perturb the kink-antikink system,
a movement of the faraway kink can induce transitions between the discrete fermion modes bound to the
solitons. These state changes can be interpreted globally as creation or destruction of a novel type
of pair: a
half-fermion and a half-antifermion. The production of the half-integral pairs will dominate over other
particle production channels as long as the solitary waves remain well separated, so that there is a
manifold of discrete fermion states whose energies are either zero or exponentially close to zero.

\bigskip

\end{titlepage}

\newpage

\section{Introduction}

Particle production by extended structures in motion is a well-studied phenomenon,
particularly in the presence of an acceleration or an acceleration-like gravitational field.
It has been analytically understood in a number of situations, including as Hawking radiation~\cite{Hawking}
around black holes and as the Unruh effect~\cite{unruh} for an accelerating dectector
(or equivalently, in terms of the radiation from moving mirrors~\cite{Davies}).
Particle production can be viewed as transitions between the quantum vacuum and Fock states containing
one or more excitations. The transitions induced by moving structures in quantum field theory are
actually analogous to situations that occur in nonrelativistic quantum mechanics,
when transitions between different states can occur even when a system is moving with constant velocity,
as shown in Refs.~\cite{Zener,1969}. Indeed, the problem of how quanta may be created
(or annihilated) by moving boundaries or structures is very rich one, with ties to problems in a number of
topics in fundamental physics; besides Hawking-Unruh radiation, there are also natural connections
to the Casimir effect~\cite{ref-jaffe1}.

In many cases of interest, the motion of the relevant boundary structures is quite slow compared to the
characteristic transit times of any real or virtual particles that may be produced.
When an eigenstate $\psi_{n}$ with energy $E_{n}$ is evolving under a time-dependent Hamiltonian $H(t)$,
there are two important time scales. One of them is the time scale describing the rate at which the
Hamiltonian is varying. The other is $\sim2\pi/|E_{n}-E_{m}|$, which is the time scale at which the phase
difference between the state $\psi_{n}$ and the nearest
other accessible eigenstate $\psi_{m}$ accumulates;
in a state that is a superposition of $\psi_{n}$ and $\psi_{m}$, this is the period over which the observables
of the system oscillate.
If the first time scale is much larger than the second, the time evolution process is called ``adiabatic'';
there are effectively
no state transitions, as the state of the system tracks the instantaneous energy eigenstate
in which it began, and the sole effect on $\psi_{n}$
due to a cyclic change in the Hamiltonian, $H(t_{1})=H(t_{0})$, will be a geometric phase
which depends on the path the Hamiltonian has taken between $t_{0}$ and $t_{1}$ and which can be
described by the Berry curvature~\cite{Berry}. Using superpositions, this kind of phase is observable;
a striking example is actually the well-known Aharonov-Bohm effect~\cite{Bohm}, where introduction of
a magnetic field in a region inaccessible to the particle beam leads to a change in the
pattern in an electron interference experiment.

In first-order perturbation theory with a time-dependent perturbation
Hamiltonian, however, one does find transitions between
unperturbed eigenstates $\psi_{n}$ and $\psi_{m}$ at a rate proportional to
$\langle m|\dot{H}|n\rangle/(E_{m}-E_{n})$. Such terms are neglected in the adiabatic approximation, meaning
that the approximation misses these weak---but potentially interesting---state transitions.
An example of this situation is treated in Ref.~\cite{1969}, which describes two plates (represented
by infinite, hard-wall potentials)
which are moving away from or towards one-another with finite velocity. In such a situation, because of the
time dependence of the Hamiltonian, the time-dependent
wave functions that satisfy the Schr\"{o}dinger equation are not
energy eigenstates, but they can be expressed in terms of the
instantaneous energy eigenstates that exist when the plates are not moving at all. The mixing of the
instantaneous eigenstates represents transitions between different states. 
Intuitively, a particular solution of the Schr\"{o}dinger equation not being an energy eigenstate tells
us that there must be energy exchanged between the plates and the quantum-mechanical particle
trapped between them. The hard wall
potential is a very special, exactly solvable system, but the argument for transitions into different states
is generic in non-adiabatic processes. In Ref.~\cite{Zener} the transition probabilities between different
energy levels was computed under a more general set of conditions; the paper considered a system of molecules
which could occupy either of two eigenstates---with either polar or nonpolar characteristics depending on
their bond lengths. As the molecules move close together and then separate, their
energy levels also get closer and then spread apart again. In such a scenario there is a finite probability
of transition between the two energy levels---a Landau-Zener transition.
To compute the probability involved, it was presumed that the difference in
energies varied approximately linearly with time, and thus with the relative velocity of a pair of molecules.
In this paper, we shall encounter a similar situation in which the difference between two bound-state
energies varies linearly with the relative velocity, although it does not
necessarily lead to any state transitions.

We can extend these idea to quantum field theories where the particle number is not conserved and
in which there may be transitions between states corresponding to different numbers of quanta. In such a
scenario, a Landau-Zener transition would not only excite a quantum particle to a higher energy but could produce
transitions between multi-particle states, which may represent either the creation or destruction of
quanta. One can try to imagine the quantum-field-theoretic extension of the hard wall potential problem, in
which each hard wall may be modeled as a very heavy solitary wave coupled to additional quantum fields,
toward which it behaves like a boundary object.
If we start from a vacuum state, with no light quanta present, and then move the one or more soliton
boundaries with some finite velocities, then there can be a nonzero
amplitude for the quantum fields coupled to the solitons to transition to single- or multi-particle states;
the soliton's motion has resulted in the creation of particles. However, particle creation is always going
to be suppressed by the differences in energy (and momentum) between the vacuum and states containing quanta.
For the creation of even a single unbound propagating particle, the energy required to effect the transition
is sizable. However, if the coupling to a solitary wave is strong enough, there may be bound states attached
to the soliton, with energies well below the continuum threshold---which may mean that 
the particle production will be dominated by creation of quanta in these bound states.
A specific illustrative model would be that of a bosonic kink or antikink coupled to a fermion field in
$1+1$ dimensions~\cite{ref-jackiw}.
The coupling between a kink and the fermion field modifies the spectrum of the fermionic
states---creating zero-energy modes, possibly other discrete states, and a modified continuum structure for
the fermions that can still escape to infinity.

In this paper, we shall be specifically studying theories with fermions
quantized in a time-dependent background composed of multiple bosonic kink structures, looking particularly
at the behavior of zero-energy and almost-zero-energy fermion states. We can approximate the
virtual fermion contributions to the vacuum energy and the pair production thresholds by essentially
neglecting the effects of the other states with energies much farther from zero. Moreover, if one or
more solitons are moving with finite velocities,
it seems that there may be some probability of pair production within
the manifold of almost-zero-energy states, due to Landau-Zener-like transitions. However, it turns out there
is actually no such effect in a system with only two solitons, as we shall show in section~\ref{sec-2moving}.
Therefore we introduce a third kink, far away from a close kink-antikink system,
and calculate the approximate fermionic energies and wave functions in this
background in section~\ref{sec-3static}. As we let the
faraway kink move with some finite velocity, we compute the probability transition between different
discrete states and interpret the result in section~\ref{sec-3moving}. Additional
details and subsidiary calculations are presented in four appendices.

\section{Fermions in a Kink-Antikink System}

\label{sec-2soliton}

Our system with scalar field solitons coupled to a fermion
field in $1+1$ dimensions is described by the Lagrange density 
\begin{eqnarray}
    \mathcal{L}= \frac{1}{2}(\partial_{\mu}\phi)^2-{
    \frac{\lambda}{4}\left(\phi^2-\frac{m^2}{\lambda}\right)\!}^{2}
    +i\bar{\Psi}\slashed{\partial}\Psi-g\bar{\Psi}\Psi\phi.
\end{eqnarray}
This $\phi^{4}$ model with a spontaneously broken discrete symmetry has a extensive history as a venue for
exploring quantum effects in theories that support classical solitary
waves~\cite{ref-jackiw,ref-dashen,ref-polyakov,ref-goldstone1,ref-rajaraman,ref-shifman1,
ref-graham1,ref-graham2,ref-goldhaber,ref-'thooft1,ref-vachaspati,ref-brihaye,ref-amado,ref-perapechka}.
As is commonly done, we shall set the scalar field parameters to be $\lambda=m^2=2$ for simplicity, so
that the Lagrange density becomes
$\mathcal{L}=\frac{1}{2}(\partial_{\mu}\phi)^2-\frac{1}{2}(\phi^2-1)^2+
\bar{\Psi}(i\slashed{\partial})\Psi-g\bar{\Psi}\Psi\phi$.
This leaves the Yukawa coupling as the principal free parameter describing the model; specifically, $g$ is
twice the ratio between the masses of the continuum fermion and boson excitations. Although we shall
derive many results that are nonperturbative in $g$, we typically assume $g\ll 1$ to simplify and make
more illustrative our analytical computations. (How some of the results in this section generalize to arbitrary
$g$ is discussed in appendix~\ref{app-review}.)

In the absence of the fermions,
the scalar field admits vacuum solutions $\phi=\pm1$ (which spontaneously break the discrete symmetry
$\phi\rightarrow-\phi$ and give the continuum fermion states mass), as well as
finite-energy kink and antikink solutions that interpolate between the two vacua.
The stationary kink solution $\phi_{K1}=\tanh{(x-x_{K1})}$ has a localized energy density centered at $x_{1}$,
with an integrated total energy $E=\frac{4}{3}$. If we assume that the fermion interactions are
weak enough that they do not destabilize the kink itself, then we can find the energy levels
(at one-loop order) for
the fermion field $\Psi$ attached to the kink by solving the Dirac equation simultaneously with
the modified scalar field equation itself. This will give us coupled differential
equations~\cite{ref-rajaraman}, which modify the energy levels of both the fermion modes and and the
nonperturbative kink. However, it turns out that for a zero-energy fermion mode, the modified field equation
for the scalar is satisfied identically, and we get an improved approximation with a minimum of
additional analytical effort.

Using the representation of the $2\times2$ Dirac matrices in which $\gamma^{0}= \sigma_{1}$ and
$\gamma^{1}= i\sigma_{3}$, the single-particle Dirac Hamiltonian is 
\begin{eqnarray}
    H&=&\alpha p +g\beta\phi \\
    &=&-i\sigma_{2}\partial_{x}+\sigma_{1}g\phi \\
    &=&\left[
\begin{array}{cc}
0 & -\partial_{x}+g\phi \\
\partial_{x}+g\phi & 0
\end{array}
\right].
\end{eqnarray}
In the background of a kink situated at $x=x_{K1}$, we can find the zero-energy fermion mode by solving
$H\Psi_{K1}=0$. The normalized solution is
\begin{eqnarray}
    \Psi_{K1} & = & \left[\frac{g\Gamma(g+\frac{1}{2})}{\sqrt{\pi}\,\Gamma(g+1)}\right]^{1/2}\left[
    \begin{array}{c}
    \sech^{g}(x-x_{K1}) \\ 0
    \end{array}
    \right] \equiv \left[
    \begin{array}{c}
    \psi_{K1} \\ 0
    \end{array}
    \right]
    \label{eq-normalization} \\
    & \approx & \sqrt{\frac{g}{4^g}}\left[
    \begin{array}{c}
    \sech^{g}(x-x_{K1}) \\ 0
    \end{array}
    \right].
\end{eqnarray}
The approximate form is valid in the $g\ll 1$ limit that we shall typically be using.
Similarly, for an antikink situated at $x=x_{A}$ the zero-energy fermion mode is given by
\begin{eqnarray}
    \Psi_{A} \equiv \left[
    \begin{array}{c}
    0 \\
    \psi_{A}
    \end{array}
    \right] & \approx & \sqrt{\frac{g}{4^g}}\left[
    \begin{array}{c}
    0 \\
    \sech^{g}(x-x_{A})
    \end{array}
    \right].
\end{eqnarray}
The presence of a zero mode in the spectrum of a quantum system indicates a degeneracy. In the presence of
a single kink or antikink, the states with the attached fermion mode filled and empty are identical in energy.
This leads to the remarkable phenomenon of fermion fractionalization~\cite{ref-jackiw}. To have a vacuum
state that is invariant under charge conjugation, the formal vacuum must contain a fractional part
of the fermion number for the zero mode;
effectively, the mode is split, half and half, between the fermion and antifermion parts of the spectrum.
The physical states, with the zero mode either fully occupied or fully unoccupied, have (after the subtraction
of the Dirac sea) fermion numbers $n_{F}=\frac{1}{2}$ and  $n_{F}=-\frac{1}{2}$, respectively.

A natural question would be what happens to the energy levels of the fermion field in presence of 
both a kink and antikink. While exact classical multiple-kink solutions are known for the closely
related sine-Gordon equation, general analytic solutions representing interacting kink-antikink pairs
are not available in the $\phi^{4}$ model. However, a smooth scalar field profile
approximating the presence of a well-separated kink and antikink is
\begin{equation}
\label{eq-phi-2sol}
\phi= \phi_{K1}+\phi_{A}-1= \tanh(x-x_{K1})-\tanh(x-x_{A})-1,
\end{equation}
which centers a kink at $x_{K1}$ and an antikink at $x_{A}$, with $x_{K1}<x_{A}$ (so that $\phi$ is
close to a vacuum value in between the solitons).
Using this expression for $\phi$,
we can approximate the fermion wave functions variationally, beginning with trial wave function of the
form $a\Psi_{K1}+b\Psi_{A}$. These describe superpositions of fermion states that are closely localized
around the kink and antikink cores. Since the (first-order, matrix) Dirac equation may be converted into
a second-order Schr\"{o}dinger-like equation for $E^{2}$, there should be no difficulties with justifying the
usual variational approach for finding the ground state of the system (even though we shall not actually
work with the second-order formulation). The correct values $a$ and $b$ should minimize the energy of the
Dirac Hamiltonian (subject to the usual normalizability condition), and in this approximation,
the ground state turns out to be 
\begin{equation}
\label{eq-Psi-}
     \Psi_{-}=\frac{1}{\sqrt{2}}\left[
    \begin{array}{c}
    \psi_{K1} \\
    \psi_{A}
    \end{array}
    \right]=\sqrt{\frac{g}{2\left(4^{g}\right)}}\left[
    \begin{array}{c}
    \sech^{g}({x-x_{K1}}) \\
    \sech^{g}({x-x_{A}})
    \end{array}
    \right],
\end{equation}
and the corresponding energy (for $g\ll 1$,
in which case the physical widths of the solitons are much smaller than the extent
of the fermion zero modes, so it is suitable to approximate $\phi$ as a sum of step functions) is
\begin{equation}
\int dx\,\Psi^{\dagger}_{-} H \Psi_{-}= - ge^{-g(x_{A}-x_{K1})}\equiv -E_{+}.
\label{eq-2sol-energy}
\end{equation}
Since this energy is negative, in the vacuum state, this mode is occupied.

The excited mode wave function
may be found by the requirement that it must be orthogonal to $\Psi_{-}$ within the
state manifold spanned by $\Psi_{K1}$ and $\Psi_{A}$. This gives $\Psi_{+}=\sigma_{3}\Psi_{-}$,
and the energy of this excited state is $E_{+}= \int dx\,\Psi^{\dagger}_{+} H \Psi_{+}= ge^{-g(x_{A}-x_{K1})}$.
$2E_{+}$ can be seen as the energy required to create a fermion-antifermion pair, by emptying the
$\Psi_{-}$ state and (to conserve fermion number) filling the $\Psi_{+}$ state.
It is also seen that the energy of the fermion field is zero when there is exactly one fermion or one
antifermion particle present. This additive nature of energy levels is characteristic of 
a (fermionic) harmonic oscillator, and it confirms our idea of a weakly interacting ground state
for $g\ll 1$.

Note that as the separation $x_{A}-x_{K1}$ grows large, these energies approach zero exponentially.
When the kink and antikink are infinitely far apart, there are two separate and degenerate zero-energy
modes---precisely what we expect for fermions coupled to two completely isolated solitary waves.
%
%
The dependence on $x_{A}-x_{K1}$ represents a spectral flow
of the energy eigenvalues away from the degenerate limiting value of $E_{\pm}=0$ that is taken when the solitary
waves are infinitely far apart. A more general variational wave function than $a\Psi_{K1}+b\Psi_{A}$ would be
expected to lead to lead to a larger (and more accurate) approximate value for the splitting between
the negative- and positive-energy bound states, and the states
would not need to be displaced by exactly equal and opposite amounts away from $E=0$.
%
%
In a time-dependent scenario in which the kink and antikink eventually approach and annihilate into
radiation, the spectral flow will also include changes in the number of discrete
fermion modes attached to the solitary waves.
As the solitons coalesce and lose their shapes, the
attractive strengths of the effective Schr\"{o}dinger potential for the upper and lower components will
grow weaker, so the additional discrete localized eigenstates that are present when $g\gtrsim1$
can flow back into the positive- and negative-energy continua.
However, in $1+1$ dimensions, the discrete states with energies closest to zero will persist as long as the
evolving scalar field profile still generates global minima of the effective potentials.

\begin{figure}
\centering
\includegraphics[width=12cm]{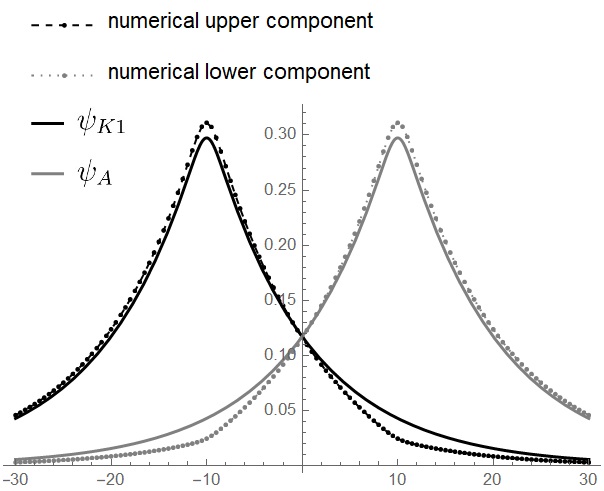}
\caption{Comparison of the upper and lower components of the variational wave function $\Psi_{-}$ and
a numerical solution of the Dirac equation, all with Yukawa coupling $g=0.1$ and soliton positions
$x_{K1}=-10$ and $x_{A}=10$. The upper components are peaked on the left, and the lower components on the
right, with the variational expressions shown as lines (black for $x_{K1}$ and gray for $x_{A}$) and
the corresponding numerically integrated values lying close by. The most notable differences between
the solutions and the variational forms are the salient points in the integrated solutions located at
$x_{A}$ for the upper component and $x_{K1}$ for the lower.
\label{fig-psi-plots}}
\end{figure}

%
Figure~\ref{fig-psi-plots} shows a comparison of the variational form $\Psi_{-}$ with the numerically
integrated solution of the Dirac equation in the background (\ref{eq-phi-2sol}), for $g=0.1$
and $g(x_{A}-x_{K1})=2$, showing good
agreement between the two, except for some small qualitative features that it is evident cannot be present
in the analytical approximation $\Psi_{-}$. In particular, there are
fairly abrupt changes in the slopes of the
upper and lower components of the numerical solution at the locations of both solitary waves, $x_{K1}$
and $x_{A}$; however, $\Psi_{-}$ only captures the peak in each fermion component at the soliton about
which is primarily localized, missing the smaller feature at the location of the farther-off soliton.
Yet that the secondary changes in slope should exist
is in fact quite clear from the form of the Dirac equation---or, equivalently, from the relation between
the upper and lower components---in the presence of the kink-antikink background (\ref{eq-phi-2sol}).

When $g<1$, each isolated one-soliton system has only a single bound state (the zero mode)~\cite{ref-dashen},
so the kink-antikink system we are considering has just the two bound fermion modes. For larger values
of $g$, this approximation gives the energies for the fermion bound
states that lie closest to zero energy (the almost-zero-energy modes).
More details about these analytic approximations for the discrete fermion states are discussed in
appendix~\ref{app-review}.

Because of the fermion energies' dependences on  on $x_{A}-x_{K1}$, the bound states
generate contribution to the effective potential between the kink and the antikink. There are
also bosonic, Casimir-like quantum forces between them~\cite{ref-graham1,ref-graham2}; however,
the character of the fermion-induced potential is somewhat different, because it depends
on which fermionic states are occupied. In the vacuum, the state $\Psi_{-}$ is occupied, while
$\Psi_{+}$ is empty. $E_{-}$ is negative and becomes more
negative as $x_{A}-x_{K1}$ decreases, creating an attraction between pair. Since
this effective interaction exists in the absence of any fermion or antifermion excitations,
it must be a product of virtual particle effects. If either a
single fermion or a single antifermion is present---that is, if both almost-zero-energy
fermion modes are either occupied or empty---the total energy vanishes, and there is no
fermion contribution to the kink-antikink effective potential at this order of approximation;
the presence of a single  quantum produces a cancellation
of the virtual particle corrections mentioned above. Finally, in the presence of both a
fermion and an antifermion ($\Psi_{+}$ occupied and $\Psi_{-}$ empty), the kink-antikink
potential term becomes repulsive.

However,
there are several alternative interpretations that might be assigned to the physics of the two
almost-zero-energy fermion states of the kink-antikink system. The ambiguity of the situation
was already noted in Ref.~\cite{ref-jackiw}, and we shall revisit it
below when we discuss the almost-zero-energy fermion modes in three-soliton backgrounds.
The crux of this issue is as follows. Globally, since there is one
such state with an actually slightly positive energy and one with a negative energy, these may be
designated according to the usual Dirac procedure, as $\Psi_{+}$ being a fermion state and $\Psi_{-}$
being a state in which a hole represents an antifermion; charge conjugation interchanges the
corresponding fermion and antifermion states. However, an alternative approximate viewpoint is possible
for an observer who is themselves ``localized'' around just one of the solitary waves.  To a very
good approximation, the discrete states still look like zero modes. To an observer only makes measurements
in the vicinity of $x_{K1}$, the system appears at short times to be just a single kink with fermion number
$\pm\frac{1}{2}$, depending on whether or not the state $\Psi_{K1}$ is filled---although this picture
needs to be adjusted at sufficiently long times. In fact, if the initial state of the full kink-antikink
system is that there is a single fermionic quantum with wave function $\Psi(t=0)=\Psi_{K1}$,
then the fermion state will evolve in time, since $\Psi_{K1}$ is not an energy eigenstate. Over time,
the quantum will tunnel back and forth with frequency $E_{+}$,
between the state $\Psi_{K1}$ localized around the kink and $\Psi_{A}$ at the antikink. From the point
of the local observer who sees only the kink, the process by which the $\Psi_{K1}$ mode evolves from
occupied to empty could be interpreted entirely in terms of the local degrees of freedom that have
half-integer fermion numbers. From this viewpoint, the time evolution looks like having
the half-fermion that is initially present tunnel away (effectively disappearing to infinity) and a
half-antifermion tunnel in from infinity to replace it. Then, after enough time has elapsed, the two half-quanta
will switch places again---and so on. This is certainly a peculiar description of what is happening, but
it appears to accord with what the local observer, looking only at the kink region, would see.

It may seem puzzling that the very meaning of the natural eigenstates of the theory may depend qualitatively
on the distance between the two solitons---that when the kink and antikink are very, very far apart, it is
logical to describe the theory in terms of local states with half-integer numbers of localized
fermions; yet when the solitons are closer together, it may be more natural to use the delocalized energy
eigenstates with integral fermion numbers. Apparently, the
appropriate assignment of states to the fermion and antifermion
parts of the spectrum depends on the parameters of the interaction background. However, this is actually
a known feature of the Dirac theory; whether particular states represent fermions or the absences of
antifermions cannot, in general, be disentangled from the background bosonic fields with which the
fermion field is interacting~\cite{ref-lieb1}.

\section{In the Moving Kink-Antikink System}

\label{sec-2moving}

Since kink-antikink pairs interact (according to the classical field equations, augmented by quantum
corrections), it makes sense to ask what happens to the fermion states when the two solitons are in
relative motion. When the solitons are in motion relative to one-another, we can always Lorentz boost to
the center of
mass frame, in which the kink and antikink are moving towards or away from each other with equal speeds.
The rate of tunneling transition in these boosted and unboosted frames will be related by a multiplicative
Lorentz factor, and hence calculation done in the center of the mass frame will suffice for our current
purposes. The solution $\Psi'_{K1'}$ for the fermion attached to a kink moving at constant velocity $v$,
centered instantaneously at $x_{K1'}=x_{K1}+vt$, can be obtained by Lorentz boosting the
previous solution for when the kink was at rest. The calculations in this section can be easily done
to all orders in the velocity $v$; however, we will be keeping terms only up to first order in $v$,
so that we can most easily see the differences between different scenarios in which calculation to all
orders in $v$ may not be so trivial (such as those we shall  encounter in section~\ref{sec-3moving}).
The wave function for the fermion, boosted with the kink's rapidity $\eta$ is 
\begin{eqnarray}
    \Psi'_{K1'}&=& e^{\frac{\eta}{2}\sigma_{2}}\left[
    \begin{array}{c}
    \sech^{g}\gamma(x-x_{K1}-vt) \\ 0
    \end{array}
    \right] \\
    &=& \left(1+\frac{\eta}{2}\sigma_{2}\right)\left[
    \begin{array}{c}
    \sech^{g}\gamma(x-x_{K1}-vt) \\ 0
    \end{array}
    \right] \\
    &=& \Psi_{K1'}+\frac{v}{2}\sigma_{2}\Psi_{K1'},
\end{eqnarray}
where we have used the $v\ll 1$ aproximations $\eta=\tanh v=v+\mathcal{O}(v^3)$ and $\gamma=1+\mathcal{O}(v^2)$;
and $\Psi_{K1'}$ is same as
$\Psi_{K1}$ except that it is centered at the moving position $x_{K1'}$. Similarly, the boosted
fermion attached to an antikink moving with velocity $-v$ is
$\Psi'_{A'}=\Psi_{A'}-\frac{v}{2}\sigma_{2}\Psi_{A'}$. We have assigned the kink and antikink equal
and opposite velocities (converging together if $v>0$), so that $\Psi'_{K1'}$ and $\Psi'_{A'}$ are the
wave functions in the kink-antikink center-of-mass frame.

In this framework, the instantaneous variational ground state is modified to
\begin{eqnarray}
    \Psi'_{-'} &=& \frac{1}{\sqrt{2}}(\Psi'_{K1'}+\Psi'_{A'}) \\
    &=& \Psi_{-'}+\frac{v}{2}\sigma_{2}\Psi_{+'},
\end{eqnarray}
where the primed subscripts $+'$ and $-'$ again refer to the use of instantaneous coordinates $x_{K1'}$
and $x_{A'}=x_{A}-vt$. The orthogonal (to first order in $v$) excited state must then be
$\Psi'_{+'}=\Psi_{+'}+\frac{v}{2}\sigma_{2}\Psi_{-'}$.
This makes the energies of the two states $E_{\pm'}=\pm(1+2gvt)E_{\pm}$. Although the energy eigenvalues
are time dependent, in the adiabatic limit we expect these states not to mix,
%
%
although there is gradual spectral flow of the eigenvalues as the separation changes.

However, as the solitons move towards each other, the wave functions which satisfy the exact Dirac equation
$(H-i\partial_{t})\Psi=0$ will really be composed of a linear combination of basis states which actually
includes more than just the moving almost-zero-energy modes---but also both any additional discrete (bound)
states that may exist and the continuum scattering states, 
\begin{equation}
    \Psi=\alpha\Psi'_{+'}+\beta\Psi'_{-'}+\text{higher discrete states}+\text{free continuum states}.
\end{equation}
Including all the states gives us various transition probabilities, which could be computed perturbatively.
However, at low velocity the dominant contribution should come only from transitions between $\Psi'_{-'}$
and $\Psi'_{+'}$, because their energies are exponentially close to zero. Evolution within this manifold
of almost-degenerate states should be faster than transitions to other states that are separated
by substantial energy gaps.
Therefore, ignoring the transitions to other states (in order to get at least a semiquantitative
analytic understanding of the behavior), the Dirac equation for the time-dependent
almost-zero-modes simplifies to
\begin{equation}
    -i\dot{\alpha}\Psi'_{+'}-i\dot{\beta}\Psi'_{-'}+\alpha(H-i\partial_{t})\Psi'_{+'}+
    \beta(H-i\partial_{t})\Psi'_{-'} =0.
\end{equation}
Note that the image of the operator $(H-i\partial_{t})$ on $\Psi'_{+'}$ or $\Psi'_{-'}$ should really
include both the almost-zero-energy wave functions and the more distant states. Although we are neglecting
these contributions to the evolving wave functions, the decomposition of the
image into these different state types does describe the coupling and transitions into higher-energy
states---paving a path for numerical or even analytical calculation in the future.
Dropping the extra terms in the decomposition for the action on $\Psi'_{+'}$,
\begin{eqnarray}
\label{eq-decomposition}
  (H-i\partial_{t})\Psi'_{+'}=E_{1}\Psi'_{+'}+E_{12} \Psi'_{-'}+\text{higher discrete}+\text{continuum},
\end{eqnarray}
the time dependence is controlled by the two parameters $E_{1}$ and $E_{12}$.

Besides these approximations, note that we have already neglected the fact that our
kink-antikink profile for $\phi$ is not an exact solution of the scalar
field equations, even in the limit in which the solitons are instantaneously stationary.
In fact, there is an extensive literature, both analytical and especially computational, on the
classical and quantum-mechanical behavior of $\phi^{4}$ theory solitary waves---how they interact at
long and short distances,
including the excitation of perturbative waves that eventually radiate away
the energy of a kink-antikink system. See for example, Ref.~\cite{ref-shnir} and references
therein for a partial overview of important results.
%
%
Specifically, it should be kept in mind that the discrete-state fermion energies are not the sole
one-loop contributors to the kink-antikink effective potential. Besides contributions from fermion continuum
states, there are also corrections coming from virtual (non-topological) excitations of the $\phi$ field.
These again can include both continuum
states, as well as discrete excitation modes tied to
individual solitary waves. The localized, discrete shape deformation modes (as well as the translational zero
mode) of an isolated kink are the bosonic analogues of the fermion bound states attached to a kink background.
Excitation of the shape deformation modes can play a key role in interactions both between kinks and
continuum scalar modes and between adjacent kinks and
antikinks~\cite{ref-izquierdo,ref-navarro-o,ref-campbell-schonfeld}.
For example, for a
sufficiently energetic colliding soliton pair, the collision may result in an inelastic rebound---in which
one or both of the solitary waves bounce back from the impact with the shape mode excited. The qualitative
effects of exciting the discrete bosonic modes in this kind of way may be quite complicated, with the long-term
behavior of a colliding kink-antikink pair displaying a ``fractal'' dependence on the initial soliton
velocities~\cite{ref-sugiyama,ref-anninos,ref-goodman1}.
A variety of approximate methods have been developed to deal with these complexities
that already exist in the purely bosonic $\phi^{4}$ theory. So although the focus of the present work is
on the effects of discrete fermion modes when the bosonic theory is Yukawa coupled to $\Psi$, it is
always important to remember that interactions between multiple solitary waves already have rich virtual
interaction structure, even before the introduction of the fermion field.

As one may guess, the parameter $E_{1}$ is equal to the instantaneous energy eigenvalue for $\Psi'_{+'}$,
$E_{1}= E_{+'}=(1+2gvt)ge^{-g(x_{A}-x_{K1})}$. This controls the evolution of the phase of the wave function.
More interesting is $E_{12}$, which will determine the rate for a fermion to make a transition from the
$\Psi'_{+'}$ state to $\Psi'_{-'}$. This can be computed directly by integrating the equation
(\ref{eq-decomposition}) after multiplying by $\Psi'_{-'}$ from the left and ignoring the contributions from
the other states---producing, up to $\mathcal{O}(v)$,
\begin{eqnarray}
    E_{12} &=& \int dx\, \Psi'^{\dag}_{-'} (H-i\partial_{t}) \Psi'_{+'} \\
     &=& \int dx\, \Psi^{\dag}_{-'}(H-i\partial_{t})\Psi_{+'}+ 
  \frac{v}{2}\left[\int dx\,\Psi^{\dag}_{-'}H\sigma_{2}\Psi_{-'}+
     \int dx\, \Psi^{\dag}_{+'}\sigma_{2}H\Psi_{+'}\right] \\
     &=& \frac{v}{2}\left[\int dx\,\Psi^{\dag}_{-'}(-i\partial_{x}+i\sigma_{3}g\phi)\Psi_{-'}+
     \int dx\, \Psi^{\dag}_{+'}(-i\partial_{x}-i\sigma_{3}g\phi)\Psi_{+'}\right] \\
     &=& \frac{v}{2}\left[\int dx\,\frac{-i}{2}\partial_{x}(\mid \Psi_{-'}\mid^2+\mid \Psi_{+'}\mid^2)+
     \int dx\, ig\phi (\Psi_{-'}^{\dag}\Psi_{+'}-\Psi_{+'}^{\dag}\Psi_{-'})\right] \\
     &=& 0.
\end{eqnarray}
A similar computation starting from the state $\Psi'_{-'}$ also naturally yields a null result for the
tunneling parameter which is responsible for producing state transitions. Hence, we actually do not get any
fermion-antifermion pair production due to the motion of the soliton background. Moreover, extending the
calculation to all orders in $v$ does not change this result. This may point towards the need to include the
more distant states that we initially neglected in our decomposition; or, more likely,
such transition simply do not
occur unless the solitons are not just moving, but accelerating. In retrospect, this might not actually
look like a surprising result, because a fermion-antifermion pair is of opposite parity compared with the
vacuum; therefore, since the perturbation is parity even, the transition rate should be zero. However, this
provides a nice consistency check, showing that this method reproduces the null result anticipated 
from parity and Lorentz symmetry arguments, and yet a generalization of this calculation will give us
non-zero amplitude when we introduce a third kink in the background; we shall discuss this interesting case
in section~\ref{sec-3moving}.

\section{A Stationary Three-Soliton System}

\label{sec-3static}

Since relative motion of the two-soliton background does not lead to any changes in the
occupation number for the almost-zero-energy fermion modes at leading order in the relative velocity,
it makes sense to look instead at more elaborate backgrounds---particularly ones that lack the reflection
symmetry of the two-soliton system. To this end,
suppose we have kink at $x_{K1}$, then an antikink at $x_{A}$, and then additionally another kink at $x_{K2}$,
with $x_{K1}<x_{A}<x_{K2}$ and a separation in distance scales: $(x_{K2}-x_{K1})\gg (x_{A}-x_{K1})$.
In our further calculations we shall denote $x_{K2}-x_{K1}$ and $x_{A}-x_{K1}$ by $\Delta_{2}$ and $\Delta_{1}$,
respectively. In the next section, when we shall be dealing with the kink at $x_{K2}$ moving towards $x_{A}$,
we use $x_{K2'}=x_{K2}-vt$ (again working only up to first order in velocity)
and similarly $\Delta'_{2}=\Delta_{2}-vt$. It can be seen that there are two separate small parameters,
$e^{-g\Delta_{1}}$ and $e^{-g\Delta_{2}}$, which are independent of each other. Our calculation will
neglect everything beyond first order of these parameters, including terms such as
$e^{-g(\Delta_{1}+\Delta_{2})}$ and $e^{-2g\Delta_{2}}$. We would like to be able to treat the
presence of the kink at $x_{K2}$ as a
perturbation to the tighter system consisting of the kink at $x_{K1}$ and the antikink at $x_{A}$.

With the introduction of the third kink, we anticipate there should be three almost-zero-energy states:
one each
with positive and negative energies and one exact zero-energy fermion state. [The existence of the exact
zero mode is guaranteed topologically
whenever the limiting values $\phi(-\infty)$ and $\phi(+\infty)$ of the scalar
background have opposite signs.]
The new positive- and negative-energy states should represent perturbed versions of
the previous $\Psi_{+}$ and $\Psi_{-}$ and states of the two-soliton system. Hence we expect the new
ground and upper excited state to be in the forms 
\begin{eqnarray}
    \Psi_{g} & = & \Psi_{-} + \frac{1}{\sqrt{2}}b_{g}\Psi_{K2}=\frac{1}{\sqrt{2}}
    \left[
    \begin{array}{c}\psi_{K1}+b_{g} \psi_{K2}\\
    \psi_{A}
    \end{array}\right] \\
    \Psi_{e} & = & \Psi_{+} + \frac{1}{\sqrt{2}}b_{e}\Psi_{K2}=
    \frac{1}{\sqrt{2}}\left[
    \begin{array}{c}\psi_{K1}+b_{e} \psi_{K2}\\
    -\psi_{A}
    \end{array}\right]\!;
\end{eqnarray}
and the zero-energy wave function may be computed exactly using the soliton profile
$\phi= \phi_{K1}+\phi_{A}+\phi_{K2}= \tanh(x-x_{K1})-\tanh(x-x_{A})+\tanh(x-x_{K2})$, revealing
\begin{eqnarray}
    \Psi_{0,\text{exact}}=\left[
    \begin{array}{c}
    \frac{\sech^{g}(x-x_{K1})\sech^{g}(x-x_{K2})}{\sech^{g}(x-x_{A})} \\
    0
    \end{array}\right]\!.
\end{eqnarray}
(Note that this wave function is not normalized.)

We can see that the zero-mode wave function is localized mostly around $x_{K2}$,
away from which its amplitude is exponentially damped---although there also is a smaller peak at $x_{K1}$.
This state is therefore mostly composed of $\Psi_{K2}$, with the other kink and the antikink acting as
perturbations. We thus expect to be able to write this zero-energy state as, at least approximately,
\begin{eqnarray}
    \Psi_{0}=\left[
    \begin{array}{c}c_{1}\psi_{K1}+c_{2}\psi_{K2} \\ 0
    \end{array}\right]\!.
\end{eqnarray}
To compute these parameters $c_{1}$ and $c_{2}$, we should simply equate
$\Psi_{0,\text{exact}}$ and $\Psi_{0}$ at the peak locations $x_{K1}$ and $x_{K2}$. 
At $x_{K2}$, equating these two and neglecting the term $c_{1}\psi_{K1}$,
as this is $\mathcal{O}\left(e^{-2g\Delta_{2}}\right)$ in this vicinity,
\begin{eqnarray}
    c_{2} & = & \frac{\sech^{g}(x_{K2}-x_{K1})}{\sech^{g}(x_{K2}-x_{A})}+
    \mathcal{O}\left(e^{-2g\Delta_{2}}\right) \\
    & \approx & e^{-g(x_{A}-x_{K1})} \\
    & = & e^{-g\Delta_{1}}.
\end{eqnarray}
Similarly, equating the two expressions for the wave function at $x_{K1}$, we get
\begin{eqnarray}
    c_{1} & = & \frac{\sech^{g}(x_{K1}-x_{K2})}{\sech^{g}(x_{K1}-x_{A})}+
    \mathcal{O}\left(e^{-g(\Delta_{1}+\Delta_{2})}\right) \\
    & \approx & e^{-g(x_{K2}-x_{A})} \\
    & = & e^{-g(\Delta_{2}-\Delta_{1})}.
\end{eqnarray}
Now, we normalize the wave function $\Psi_{0}$ and the final result, to the necessary order, is 
\begin{equation}
    \Psi_{0}=\left[
    \begin{array}{c}
    c\psi_{K1} +\psi_{K2} \\ 0
    \end{array}
    \right]\!,
\end{equation}
where $c=c_{1}/c_{2}=e^{-g(\Delta_{2}-2\Delta_{1})}$.

Finally, requiring that the discrete-state wave functions $\Psi_{g}$, $\Psi_{e}$, and $\Psi_{0}$ 
all be mutually orthogonal fixes the remaining parameters to be $b_{g}=b_{e}\equiv b = -(c+d)$,
where,
\begin{equation}
d\equiv\int dx\, \psi_{K1}^{*}\psi_{K2}=(1+g\Delta_{2})e^{-g\Delta_{2}}
\approx g\Delta_{2}e^{-g\Delta_{2}},
\end{equation}
using that the step function approximation for the soliton profiles that we have employed in our
calculations entail that we are in the $g\Delta_{1}$, $g\Delta_{2}\gg1$ regime.
Without taking the motion of the solitons into account, these are not only orthogonal but
also stationary states---since for these eigenfunctions of a time-independent Hamiltonian,
we must automatically have $\int dx\, \Psi_{\xi}^{\dag}H\Psi_{\zeta}= 0$
(where $\xi$ and $\zeta$ may be $e$, $g$ or $0$),
unless $\xi=\zeta$. For the diagonal matrix elements (the energy
expectation values), we have
\begin{equation}
\int dx\, \Psi_{\xi}^{\dag}H\Psi_{\xi}=\left\{
\begin{array}{ll}
-ge^{-g\Delta_{1}}, & \xi=g \\
0, & \xi=0 \\
ge^{-g\Delta_{1}}, & \xi=e
\end{array}
\right.\!\!.
\end{equation}
Further details of the calculations are given in appendix~\ref{app-3stationary}.

Now, circling back to our discussion from the end of section~\ref{sec-2soliton}, we shall try to
understand what a local observer situated near $x_{K1}$ would witness in this
situation as the fermion modes' time evolution proceeds.
To understand this, note that evolution of $\Psi_{K1}$ is given by
\begin{equation}
\label{eq-PsiK1-local}
\Psi_{K1}(t)= \frac{1}{\sqrt{2}}\left(\Psi_{g}e^{iE_{+}t}+\Psi_{e}e^{-iE_{+}t}\right)-b\Psi_{0}.
\end{equation}
It is easy to see that as time passes, there is again the
anticorrelated tunneling of a half-fermion and a half-antifermion at $x_{K1}$,
just as discussed at the end of section~\ref{sec-2soliton}. More importantly, for
localized observations around $x_{K1}$ the additional term $b\Psi_{0}$ will bring about no
significant change---since the wave function $\Psi_{0}$ appearing in this term
term is $\mathcal{O}(e^{-2g\Delta_{2}})$ at $x_{K1}$. So locally, the observer still sees tunneling of
half-quanta with frequency $E_{+}$, even after a
stationary third soliton is introduced very far away.

\section{When the Faraway Kink is Moving}
 
\label{sec-3moving}

To find a system in which there are nontrivial fermion state transitions, we may
let the more distant kink at $x_{K2}$ now approach (moving with velocity $v$)
the more compactly spaced kink-antikink pair. We will again perform calculations up to first order in
this velocity.
It is now convenient to define new co-ordinate $x_{K2'}= x_{K2}-vt$. Our new basis states are $\Psi_{\mp}$
and $\Psi'_{0}$, where $\Psi'_{0}$ is the Lorentz-boosted fermion state given by
\begin{eqnarray}
     \Psi'_{0}&= &e^{-\frac{\eta}{2}\sigma_{2}}\Psi_{K2'} \\
              & \approx & \left(1-\frac{v}{2}\sigma_{2}\right)\Psi_{K2'}.
\end{eqnarray}
We want to identify the approximate states that instantaneously represent the ground, excited, and
zero-energy states of the system. For exact zero modes (and when the charge conjugation symmetry C is not
broken~\cite{ref-goldstone-wilczek}), we must still have the Jackiw-Rebbi fermion fractionalization;
such states may be interpreted as representing a net
$n_{F}=\frac{1}{2}$ fermion
present when they are occupied, versus $n_{F}=-\frac{1}{2}$ when they are unoccupied. This happens
as a result of the renormalization subtraction of the infinite number of fermions in the negative-energy
Dirac sea. In order to perform this subtraction in a C-invariant fashion, a zero-energy state must
be assigned half to the set of negative-energy sea states and half to the set of positive-energy states.
As a consequence, whenever an odd number of solitons are present---so that the vacuum configurations at
$x=-\infty$ and $x=+\infty$ are different---the fermion number eigenvalues will be half-integers.

Now, it is natural to expect that instantaneously the almost-zero-energy states
should be represented by modifying the previous stationary states from depending on $\Psi_{K2}$ 
to depending on the boosted $(1-\frac{v}{2}\sigma_{2})\Psi_{K2'}$ instead. This intuition is further
supported by a calculation of the energies of these modified states, which
match exactly with our previous results; the details are given in appendix~\ref{app-3moving}.
So our modified states are
\begin{eqnarray}
    \label{eq-Psiprime-e}
    \Psi'_{e} & = & \Psi_{e}- \frac{b'v}{2\sqrt{2}}\sigma_{2}\Psi_{K2'} \\
    \label{eq-Psiprime-g}
    \Psi'_{g} & = & \Psi_{g}- \frac{b'v}{2\sqrt{2}}\sigma_{2}\Psi_{K2'} \\
    \Psi'_{0} & = & \Psi_{0}- \frac{v}{2}\sigma_{2}\Psi_{K2'}.
\end{eqnarray}
It is understood that these depend
on revised parameters parameters $b'$ and $c'$, each of which depends on the the
instantaneous position of the faraway kink $x_{K2}-vt$ and the instantaneous separation
$\Delta_{2}'=x_{K2'}-x_{K1}$ in the same way that $b$ and $c$ depended on $\Delta_{2}$.
This dependence enters both explicitly in the formulas (\ref{eq-Psiprime-e}) and (\ref{eq-Psiprime-g}),
as well as implicitly---since the wave functions $\Psi_{e}$ and $\Psi_{g}$
include a $\Psi_{K2}$ part and thus depend upon the instantaneous position of the moving kink.
The wave functions $\Psi'_{e}$ and $\Psi'_{g}$ are no longer orthogonal to $\Psi'_{0}$ on their own,
and their values are
given in the appendix. It might seem that the non-orthogonality might be avoided if one simply got rid of
the $-\frac{v}{2}\sigma_{2}$ term in the boost modification with
$\Psi_{K2'}$, keeping only the change in coordinates
$x_{K2'}= x_{K2}-vt$, which naively seems like a nonrelativistic change, whereas the $-\frac{v}{2}\sigma_{2}$
term seemingly represents a relativistic effect coming from the boost. However, the only way the $x_{K2'}$
differs from $x_{K2}$ is through the effects of its time derivatives, which would always be associated with
factors of the inverse speed of light (were that maintained as a dimensional parameter);
hence they both represent relativistic corrections of the same
order and thus must be included together. We also demonstrate in appendix~\ref{app-3moving}
that the occupation numbers of these states evolve in time, with a ``tunneling'' parameter
\begin{equation}
\lambda_{2}\equiv
\int dx\, \Psi'_{0}H \Psi'_{e}= \int dx\, \Psi'_{0}H \Psi'_{g}=
-\frac{ig^2v}{\sqrt{2}}(\Delta_{2}'-\Delta_{1})e^{-g(\Delta_{2}'-\Delta_{1})}.
\end{equation}

Now, as the distant kink moves, let $\Psi$ be a solution of Dirac equation.
Resolving this into ground, excited, and zero-energy states,
\begin{eqnarray}
    \Psi= \alpha \Psi'_{e}+\beta\Psi'_{0}+\gamma\Psi'_{g},
\end{eqnarray}
the Dirac equation gives
\begin{eqnarray}
  0 & = &  (H-i\partial_{t})\Psi \\
  \label{eq-Dirac-abc}
  &=& -i\dot{\alpha}\Psi'_{e}+\alpha(H-i\partial_{t})\Psi'_{e}-i\dot{\beta}\Psi'_{0}+
  \beta(H-i\partial_{t})\Psi'_{0}-i\dot{\gamma}\Psi'_{g}+\gamma(H-i\partial_{t})\Psi'_{g} \\
  &=& (-i\dot{\alpha}+E_{+}\alpha+\tau\beta)\Psi'_{e}+
  \left[-i\dot{\beta}-(\alpha+\gamma)\tau\right]\Psi'_{0}+(-i\dot{\gamma}-E_{+}\gamma+\tau\beta)\Psi'_{g}.
\end{eqnarray}
We have used results from appendix~\ref{app-3moving} to get these final equations.
The parameter $\tau$ is derived in the appendix and has the value
$\tau=-\frac{igv}{\sqrt{2}}e^{-g(\Delta_{2}'-2\Delta_{1})}$. The key fact about $\tau$ is
that it is $\mathcal{O}(v)$. Keeping terms up to this order, it can be shown that
\begin{eqnarray}
\label{eq-final0}
 -i\dot{\alpha}+E_{+}\alpha+\tau\beta=0 \\
 \label{eq-final0b}
 -i\dot{\beta}-\tau(\alpha+\gamma)=0 \\
 \label{eq-final0c}
 -i\dot{\gamma}-E_{+}\gamma+\tau\beta=0
\end{eqnarray}
If we are interested in the problem of a fermionic quantum that begins primarily in the
instantaneous zero-mode (largely localized around $x_{K2}$), but which may
subsequently tunnel into the other states,
we may take the the initial value of the zero-mode amplitude $\beta$ to be a $\beta_{0}$, compared to
which the other initial values $\alpha_{0}$ and $\gamma_{0}$ may be small.
We can now directly integrate (\ref{eq-final0}--\ref{eq-final0c}) to obtain,
\begin{eqnarray}
\label{eq-final1}
    \alpha(t) & \approx & \alpha_{0}e^{-iE_{+}t}-\frac{\beta_{0}\tau}{E_{+}}\left(1-e^{-iE_{+}t}\right)\\
    \label{eq-final2}
    \gamma(t) & \approx & \gamma_{0}e^{iE_{+}t}-\frac{\beta_{0}\tau}{E_{+}}\left(1-e^{iE_{+}t}\right).
\end{eqnarray}
(Further details of this calculation are shown in appendix~\ref{app-ODE}.)
This solution, given this kind of initial configuration, should be valid so long as $t\ll e^{g\Delta_{2}'}$.
However, if $t$ is too great, then naturally we will need a higher-order calculation.

To see how the quanta move from state to state, consider a configuration  in which initially only
$\Psi'_{0}$ is occupied: $\beta_{0}=1$ , $\alpha_{0}=0$ and $\gamma_{0}=0$. The total fermion number in this
state is $-\frac{1}{2}$. Under the global interpretation, assigning fermion versus antifermion identity to
the delocalized instantaneous energy eigenstates, this fermion content may be interpreted
as consisting of one whole antifermion (corresponding to the unoccupied negative-energy mode
$\Psi_{g}'$) and half a fermion (from the occupied $\Psi_{0}'$).
It can be seen from (\ref{eq-final1}--\ref{eq-final2}) that the time evolution gives rise to nonzero
occupation amplitudes $\alpha$ and $\gamma$, indicating a reshuffling of the fermion states. A state in which
solely $\Psi'_{e}$ is occupied also has fermion number $-\frac{1}{2}$, but it is made up of one fermion
(occupied $\Psi'_{e}$), a half antifermion (unoccupied $\Psi'_{0}$) and one antifermion
(the empty $\Psi'_{g}$).
Finally, when $\Psi'_{g}$ is occupied, the state is made up of just half an antifermion
(since $\Psi'_{0}$ is unoccupied) and thus possesses the same fermion number $-\frac{1}{2}$.

The fermion number is necessarily
conserved when the single occupying Dirac quantum tunnels between states---from the initial zero-energy state
to either the excited or the ground state. However, when it transitions to filling the excited state, this can
be interpreted
(globally) as the creation of  production of a novel pair: a half fermion and a half-antifermion. Likewise,
when it transitions to filling the negative-energy ground state, then there is destruction of
half-fermion-plus-half-antifermion pair.
This is an interesting new structure for particle production through soliton interactions. In the Dirac sector,
the excitations produced by the relative motions of the kinks
are---according to this interpretation---dominated by creation and annihilation of bound half-particle pairs.

However, it is also interesting to consider whether there is any change in the picture seen by
the local observer at  $x_{K1}$ who observed the simultaneous tunneling of a
half-fermion and half-antifermion each back and forth between the local kink and the nearby antikink
(as discussed at the end of sections~\ref{sec-2soliton} and~\ref{sec-3static}). To address this,
we note that if at $t=0$ the local observer sees the fermion mode attached to the kink at $x_{K1}$
filled, this wave function has the decomposition
$\Psi_{K1}=\frac{1}{\sqrt{2}}(\Psi'_{e}+\Psi'_{g})-b\Psi'_{0}$.
However, the time dependence of $\Psi_{K1}$ is now more complicated than (\ref{eq-PsiK1-local}).
The occupation amplitudes evolve according to (\ref{eq-final1}--\ref{eq-final2}), with the
initial condition $\alpha_{0}= \gamma_{0}= \frac{1}{\sqrt{2}}$ and $\beta_{0}= -b$. Solving this to
the order to which we have been working, the coefficient of $\Psi'_{e}$ becomes 
$\alpha=\frac{1}{\sqrt{2}}e^{-iE_{+}t}-{\cal O}(e^{-2g\Delta_{2}})\approx\frac{1}{\sqrt{2}}e^{-iE_{+}t}$,
which is still effectively the same result as in sections~\ref{sec-2soliton} and~\ref{sec-3static}.
So if the single occupying fermion is initially localized around $x_{K1}$, the local observer
sees minimal change to the oscillatory behavior. However, we also note that in this particular case,
(\ref{eq-final1}--\ref{eq-final2}) do not show any evidence of transitions between $\Psi_{K1}$ and
$\Psi_{A}$, hence the unchanged result is perhaps not so surprising. On the other hand, we know that if
$\beta_{0}=1$ and $\alpha_{0}=\gamma_{0}=0$, then there is a local influx of half-fermion density
to $x_{K1}$ and $x_{A}$.
With these initial conditions, the local observer at $x_{K1}$ sees that there is some amplitude
representing tunneling of a half fermion toward $x_{K1},$ but it is suppressed by a factor proportional to
$ve^{-g\Delta_{2}}$ and oscillates with a frequency $E_{+}$.
More precisely, the coefficient of $\Psi_{K1}$ as the time evolves is proportional to
$-\frac{\sqrt{2}\tau}{E_{+}}(1-\cos{E_{+}t})+b$.
Note that $\tau$ is imaginary, whereas $b$ is real. The imaginary contribution may be interpreted as
coming from the production of a pair with a half-fermion and half-antifermion.
This exotic pair production leads to a (nonzero but exponentially suppressed)
tunneling amplitude for the half-fermion over to the vicinity of $x_{K1}$. 

We previously mentioned the potential similarity between this kind of pair production and a
Landau-Zener transition. However, the amplitudes we have found cannot be directly connected to the
Landau-Zener formula---which says that for two states $\phi_{1}$ and $\phi_{2}$,
with energies $E_{1}$ and $E_{2}$ that approach
very close together, the probability of the adiabatic approximation being violated by a transition
is $1-e^{-2\pi\Gamma}$, where $\Gamma=2\pi H_{12}^2/[\partial_{t}(E_{1}-E_{2})]$, in
terms of the off-diagonal matrix element of the Hamiltonian $H_{12}$. In the limit in which $H_{12}$ is
small, the non-adiabatic transition probability is approximately
$2\pi\Gamma$, which differs in form from the squares of the transition terms
in~(\ref{eq-final1}--\ref{eq-final2}), which are proportional to $v^{2}$. Although there is a common
exponential suppression of the transition rate ($\tau$ being an exponentially small function of the
system parameters), there are also several reasons apparent for this dissimilarity.
The difference in energies is not strongly-enough
time dependent to our order of calculation (as shown in
appendix~\ref{app-3moving}); thus there is no avoided crossing of two levels that have otherwise
stable energies in the $t\rightarrow\pm\infty$ limits (unless perhaps a kink and an
anti-kink actually collide; but that is a situation well outside the limits of our approximation regime).
In fact, the Landau-Zener formula assumes that we are effectively studying
the evolution of a system from the ultimate past to the ultimate future, which for the present system
under consideration would necessitate the inclusion of higher-order terms that we have uniformly neglected.
Nevertheless, the system of solitons and attached fermion modes
does have all the key elements which come together produce Landau-Zener transitions in other situations.
The off-diagonal Hamiltonian terms are present, and the differences in energy levels do vary in time, as
shown in section~\ref{sec-2moving}. In fact, at intermediate times, the solutions of the coupled differential
equations~\cite{Zener} for the coefficients $C_{1}$ and $C_{2}$ of the two aforementioned
states $\phi_{1}$ and $\phi_{2}$ can, with a similar choice of initial conditions, look exactly
like~(\ref{eq-final1}--\ref{eq-final2}). In short,
the transition amplitudes we have computed are physically similar to those in the Landau-Zener problem,
but integrated only up through an intermediate time scale.

It is also natural to ask what the
charge-conjugate analogue of (\ref{eq-final1}--\ref{eq-final2}) would be.
To answer this, we shall  need to construct wave functions analogous to $\Psi_{e}$, $\Psi_{g}$,
and $\Psi_{0}$, but for which the global fermion number is $n_{F}=+\frac{1}{2}$. These are states for
which exactly two of the fermion modes localized around $x_{K1}$, $x_{A}$, and $x_{K2}$ are
occupied; they are created by the action of two fermion creation operators upon the vacuum-like state in
which all three of the almost-zero-energy modes are empty.
We denote these new wave functions by $\Psi_{Ce}$, $\Psi_{C0}$, and $\Psi_{Cg}$. Expressed
as antisymmetrized two-particle wave functions for the two fermions that the operators create, they
take the forms
\begin{eqnarray}
\Psi_{Ce}(x_{1},x_{2})&=& \frac{1}{\sqrt{2}}[\Psi_{g}(x_{1})\otimes\Psi_{0}(x_{2})
-\Psi_{0}(x_{1})\otimes\Psi_{g}(x_{2})]\nonumber\\
\Psi_{C0}(x_{1},x_{2})&=& \frac{1}{\sqrt{2}}[\Psi_{g}(x_{1})\otimes\Psi_{e}(x_{2})
-\Psi_{e}(x_{1})\otimes\Psi_{g}(x_{2})]\nonumber\\
\Psi_{Cg}(x_{1},x_{2})&=& \frac{1}{\sqrt{2}}[\Psi_{0}(x_{1})\otimes\Psi_{e}(x_{2}).
-\Psi_{e}(x_{1})\otimes\Psi_{0}(x_{2})].
\end{eqnarray}

The Hamiltonian anticommutes with charge conjugation, and since $H\Psi_{e}=E_{+}\Psi_{e}$, the\
corresponding charge conjugated version of this equation ought to be $H\Psi_{Ce}=-E_{+}\Psi_{Ce}$.
Using the decomposition $H= H_{1}\otimes I+I\otimes H_{2}$
of the Hamiltonian into single-particle Dirac operators,
it can be straightforwardly checked that it is indeed true.
When the third kink at $x_{K2}$ is in motion, then our two-particle wave functions become $\Psi_{Ce}'$,
$\Psi_{C0}'$, and $\Psi_{Cg}'$, analogously to the moving wave functions in the $n_{F}=-\frac{1}{2}$
sector. We let $\Psi$ be the wave function which solves the two-particle Dirac equation; supposing that
only the almost-zero-energy modes are appreciably occupied, 
$\Psi$ admits the decomposition,
\begin{eqnarray}
    \Psi= \alpha_{C}\Psi_{Ce}'+\beta_{C}\Psi_{C0}'+\gamma_{C}\Psi_{Cg}'.
\end{eqnarray}
Acting on this with $(H-i\partial_{t})\Psi=0$ should give us three equations homologous to
(\ref{eq-final0}--\ref{eq-final0c}).
In fact, a little bit of algebra demonstrates that the differential equations for the time evolution of
$\alpha_{C}$, $\beta_{C}$, and $\gamma_{C}$ are identical with those for $\alpha$, $\beta$, and $\gamma$,
except for the inversion of the Hamiltonian matrix elements, $E_{+}\rightarrow E_{-}=-E_{+}$
and $\tau\rightarrow-\tau$, providing a clear demonstration
of the C invariance. As the kink at $x_{K2'}$ approaches the other solitons (in whose vicinity the
net number of fermion vanishes), the behavior of the system
is essentially the same whether the faraway kink is carrying a half-fermion or a half-antifermion. Either
half-quantum may tunnel over to the close kink-antikink region.

\section{Conclusions and Discussion}

\label{sec-concl}

We have already stated that for the localized observer in the vicinity of $x_{K1}$, (\ref{eq-final1}) and
(\ref{eq-final2}) predict that with an initial condition  of $\Psi'_{0}$ being occupied
(i.e. $\beta=\beta_{0}$ and $\alpha_{0}=\gamma_{0}=0$), there is a tunneling amplitude for a
half-fermion to arrive, and the tunneling rate also oscillates with frequency $E_{+}$. It is fruitful,
however, to discuss the
different interpretations of this situation in terms of what different observers (operating at different
spatiotemporal scales) would see, and to see how all their observations can be reconciled. In that spirit,
consider another observer who is making partially localized measurements in the vicinity of the close
kink-antikink pair; this observer can see quanta interacting over length scales of
$\mathcal{O}(\Delta_{1})$. In the the aforementioned scenario, what this observer sees does not require an
understanding of half-fermion pairs---in contrast to the situation for the previous observer who was
completely localized around $x_{K1}$. The observer monitoring the whole kink-antikink region can describe
any transitions in terms of two energy states equally displaced around zero. If this observer were to include
the effect of  all the other discrete and scattering states that we have neglected so far,
then the observer would see a positive tower of states above $\Psi_{e}$ and negative tower of states below
$\Psi_{g}$, but obviously no zero-energy state. The lack of a strictly-zero-energy mode means that,
according to what this observer sees within their regional bailiwick,
there should be no need to posit the existence of half-integer fermion numbers.
When the zero energy state $\Psi'_{0}$ is initially occupied, this observer initially sees one antifermion
in the observation region (because of the essentilly unoccupied $\Psi'_{g}$),
and the energy of the fermion field around the close kink-antikink system is consequently zero.
There is neither attraction nor repulsion between these solitary waves due to the fermionic fields.
If, after some
time has passed, $\Psi'_{e}$ becomes occupied, the kink-antikink observer sees this as one fermion
having tunneled in from infinity, bringing with it enough energy to lift the total to from zero to $E_{+}$.
This local observer can furthermore see that now the close kink-antikink pair repel each other. On the other
hand if $\Psi'_{g}$ becomes occupied (which is equally likely) then the local observer sees that the
existing antifermion has tunneled away from the region toward spatial infinity,
and now the kink-antikink pair attract one-another due to virtual particle effects. In short, this observer
either sees tunneling of a whole fermion towards the region, resulting in a fermion-antifermion pair; or
the initially antifermion tunneling away, leaving the region in its local vacuum configuration.

In contrast, from the viewpoint of another local observer positioned at $x_{K2'}$, there is initially a
half-fermion present. To this observer, regardless of whether $\Psi'_{e}$ or $\Psi'_{g}$ becomes occupied,
what is left behind at $x_{K2'}$ is a half-antifermion.
For this local observer, the half-fermion has tunneled away and a half-antifermion has tunneled inwards
to replace it.

Finally, we may consider a
fully delocalized observer who can see what is happening around
all the solitary waves and is capable of observing delocalized states
above $\mathcal{O}(\Delta_{2})$ in size.
Since the background scalar field $\phi$  takes different vacuum expectation values at $+\infty$ and
$-\infty$, this observer will know of the existence of a single exact zero mode, necessitating
the existence of half-integral fermion numbers to describe the relevant physics.
If this observer tried to write down a quantum field theory describing the quasiparticles
of this theory, then due to the renormalization subtraction of the infinite number of fermions in the
negative-energy Dirac sea, the observer would end up positing the existence of half-integral fermion numbers.
This broadest observer must be capable of reconciling all the different interpretations seen by
all the more localized observers we have talked about so far.
For this delocalized observer, the occupation of $\Psi'_{e}$ will represent tunneling of one full fermion,
but the observer must account for the fact that there was only a single half-fermion localised around
$x_{K2'}$  to begin with. Therefore this observer may conclude that there has been creation of a
half-fermion-and-half-antifermion pair. The half-antifermion appears at $x_{K2'}$, while the half-fermion 
seemingly ``merges'' with the half-femrion that was initially located at $x_{K2'}$ to become the full
fermion that moves over to the close kink-antikink region.
Similarly, when the time evolution leads to $\Psi'_{g}$ being occupied,
the initial antifermion localized around the close kink-antikink pair tunnels to the faraway kink at $x_{K2'}$,
and there half of it ``annihilates'' the half-fermion present, leaving the same net half-antifermion at
$x_{K2'}$ as in the previous case.
In each of these transitions, a quantity of energy of magnitude $E_{+}$ has been either added to or
taken away from the fermion fields, depending precisely on whether the interpretation calls for the
creation or destruction of a half-fermion-and-half-antifermion pair! The ultimate source or sink for this
energy must be the motion of the solitary wave at $x_{K2'}$. 

We have computed the probability of production or destruction of exotic pairs that may be interpreted
as consisting of one half-fermion and one half-antifermion each. Moreover, this kind of
qualitative behavior seems like it should be ubiquitous in ($1+1$)-dimensional
systems containing odd numbers of
$\phi^{4}$ solitons. Yet there is clearly plenty of room for improved understanding of
these kinds of processes. For example, our calculations have
entirely neglected the quantitative impact of higher-energy
fermion and antifermion states, including the infinite collection of continuum states. These kinds of
approximation cause problems at longer times. If the initial fermion wave function is a superposition of the
ground and excited states, we do not find any creation of fermion-antifermion pairs; however,
the total probability is not conserved, because we did not use a complete basis when
expanding the image of the operator $(H-i\partial_{t})$ acting on the discrete eigenstates.

This shows that
the multi-particle spectrum must be included if we are to understand the time evolution of such
states fully. So performing analogous calculation with the inclusion of multi-particle states
would be a natural way to extend this work. On the other hand, it would also be interesting to study whether
the creation of pairs with full fermions
and antifermions might actually take place, by extending our calculations to higher
orders or by including additional solitons---while still working entirely within the manifold of
almost-zero-energy states. Another issue that we have not yet addressed is related to the fact that
that when the soliton at $x_{K2}$ is set in motion, we have identified $\Psi'_{\pm'}$ and $\Psi'_{0}$ as
the states with the same matter content as the states $\Psi_{\pm}$ and $\Psi_{0}$ of the stationary system.
When the solitary waves are very far apart, so that the each individual wave functions do no significantly
overlap, this is should be a valid approximation. For perturbative calculations, in which fermionic
mode energies can simply be added up to find how the overall energy of the multi-soliton system is affected,
this kind of identification makes logical sense. However, it is unclear what kinds of corrections we should
expect---or if such identification is at all possible---when the solitary waves approach each other more
closely, so that the wave functions overlap significantly, potentially
giving rise to nontrivial interaction and exchange energies.

Since fractionalized states may also exist for configurations with fermions coupled to topologically nontrivial
bosonic spin-0 and
spin-1 fields in $3+1$ dimensions (including monopoles~\cite{ref-polyakov,ref-'thooft2} and
dyons~\cite{ref-julia}), it may
be interesting to expand the kind of analysis performed in this paper to such systems as well. However,
with more than one spatial dimension, things obviously become quite a bit more complicated. There are more
possibilities for how three (or more) solitons may be positioned in space,
and the spin structure of the fermion field
adds another complicating factor. The presence of the spin quantum number in $3+1$ dimensions introduces
additional
potential degeneracy; depending on precisely how the solitonic gauge fields are coupled to the Dirac field,
it seems that there may or may not be multiple exact zero-energy modes in the presence of two or more
solitary wave structures~\cite{ref-altschul42}.
In fact, there is probably quite a bit of information left to be teased out about
these systems, and addressing these various additional questions, whether in $1+1$ or $3+1$ dimensions, may
provide deeper insights into the dynamics of systems with
fermions, spontaneous symmetry breaking, and topologically nontrivial bosonic field
configurations more generally.

\section*{Acknowledgments}

B. A. is grateful to the late R. Jackiw for many years of helpful discussions and, in particular, introducing
him to the topics of solitary waves and fermion fractionalization. The authors also appreciate the
assistance of A. Rout with the numerical solution of the bound-state Dirac equation.

\appendix

\section{Fermion Energies for the Two-Soliton System}

\label{app-review}

In this appendix, 
we shall evaluate the energy expectation values $\int dx\,\Psi_{\pm}^{\dag}H\Psi_{\pm}$ for the
symmetric and antisymmetric variational wave functions, first under the approximations $g\ll 1$
and $(x_{A}-x_{K1})\gg 1$, for which elementary expressions
can be obtained explicitly. Then we shall present an analysis for general
$g$, which reduces to the previous when the propagating fermion mass is much smaller than the
propagating boson mass.
Many of these results were previously laid out in Ref.~\cite{ref-altschul42}.

When a kink and antikink are far apart, we may expect the potential due to one soliton
to be a small perturbation of the Dirac equation for the bound state that is
tightly localized around the other soliton.
In this spirit, we shall calculate the expectation values of the energies for the
symmetrized states $\Psi_{\mp}$. As an ingredient in this, we have, for example, for $H\Psi_{K1}$,
\begin{eqnarray}
H\Psi_{K1} & = & \left(-i\sigma_{2}\partial_{x}+\sigma_{1}g\phi_{K1}\right)\Psi_{K1}+
\sigma_{1}g\left(\phi_{A}-1\right)\Psi_{K1} \\
\label{eq-H+exact}
& = & 0 + \sigma_{1}g(\phi_{A}-1)\Psi_{K1}.
\end{eqnarray}
To find the perturbed energies, we evidently need to find the spatial integrals of
quantities such as $(\Psi_{+}^{\dag}\sigma_{1}\Psi_{-})
\phi_{A}$. The characteristic spatial extent of the soliton solution $\phi_{A}$ is
$\sqrt{\lambda}/m(=1)$, while the characteristic spatial decay length of the zero-mode fermion
localized near $x_{A}$ is $g^{-1}$. If $g\ll 1$, the fermion wave functions decays
very little over the actual width of the antikink, making it a good
approximation to replace $\phi_{A}$ by the signum function $-\sgn(x-x_{A})$ in the integral.
Approximating $\phi_{A}$ with this step function gives
  \begin{equation}
\label{eq-H+}
H\Psi_{K1}=-\sigma_{1}g[\sgn(x-x_{A})+1]\Psi_{K1},
\end{equation}
with the corresponding expression for $H\Psi_{A}$ being 
\begin{equation}
\label{eq-H-}
H\Psi_{A}=\sigma_{1}g[\sgn(x-x_{K1})-1]\Psi_{A}.
\end{equation}

Combining (\ref{eq-H+}) and (\ref{eq-H-}), we find
\begin{equation}
H\Psi_{\mp}=\frac{1}{\sqrt{2}}\sigma_{1}g\left\{-[\sgn(x-x_{A})+1]\Psi_{K1}
\pm[\sgn(x-x_{K1})-1]\Psi_{A}\right\}.
\end{equation}
Since $\Psi_{K1}^{\dag}\sigma_{1}\Psi_{K1}=\Psi_{A}^{\dag}\sigma_{1}\Psi_{A}=0$,
only the cross terms in $\Psi_{\pm}^{\dag}H\Psi_{\pm}$ are nonzero (even without
the spatial integration). Using the relations
\begin{equation}
\Psi_{K1}^{\dag}\sigma_{1}\Psi_{A}=\Psi_{A}^{\dag}\sigma_{1}\Psi_{K1}=\frac{g}{4^{g}}
\sech^{g}(x-x_{K1})\sech^{g}(x-x_{A}),
\end{equation}
the expectation integrals for for $E_{\mp}$ are
\begin{eqnarray}
E_{\mp} & = & \mp \frac{g^{2}}{2\left(4^{g}\right)}\int dx\,
\sech^{g}\left(x-y+\frac{\Delta}{2}\right)\sech^{g}\left(x-y-\frac{\Delta}{2}\right) \nonumber\\
& & \times\left[2-\sgn\left(x-y+\frac{\Delta}{2}\right)+\sgn\left(x
-y-\frac{\Delta}{2}\right)\right]\!,
\label{eq-sgn-integral}
\end{eqnarray}
where $y\equiv\frac{x_{K1}+x_{A}}{2}$ is the center of mass
location, and $\Delta\equiv x_{A}-x_{K1}$ is the
separation between the solitary waves. Shifting he integration $x\rightarrow(x-y)$
clearly gives a formula that depends only on $\Delta$. Moreover, the formula shows that
the zero-energy eigenstates of the one-soliton backgrounds infinitely far apart bifurcate  
into two states that are automatically symmetrically displaced above and below $E=0$
as the solitons approach one-another. To proceed further with the $g\ll 1$ estimate of
$E_{\mp}$, we insert the forms for the fermion wave functions that
correspond to the step function approximation for the hyperbolic tangent scalar profile,
\begin{equation}
\label{eq-coshapprox}
\sech(x-x_{0})=2\left[\theta(x-x_{0})
e^{(x-x_{0})}+\theta(x_{0}-x)e^{-(x-x_{0})}\right]^{-1}.
\end{equation}
Using (\ref{eq-coshapprox}), $E_{\pm}$ becomes just an integral over exponential and
step functions, which yields
\begin{equation}
E_{\mp}=\mp ge^{-g\Delta},
\end{equation}
as in (\ref{eq-2sol-energy}). An interaction energy of this nature was already suggested in
the original paper on fermion fractionalization via soliton interactions~\cite{ref-jackiw},
although that the estimate in that paper missed what we shall discuss next:
the more complicated dependence of the fermion energies on $g$ when $g$ is not necessarily small.

In fact, the result (\ref{eq-2sol-energy}) is actually just the leading term
in the fermionic interaction energy when $g^{-1}$ is large.
From the derivation of (\ref{eq-sgn-integral}) it should be clear that dropping the $g\ll 1$
approximation only requires replacing the signum functions with the more accurate
hyperbolic tangent profiles for the kink and antikink and using the more precise
wave function normalization constant from (\ref{eq-normalization}),
\begin{eqnarray}
E_{\mp} & = & \mp \frac{g^{2}\Gamma(g+\frac{1}{2})}{2\sqrt{\pi}\Gamma(g+1)}\int dx\,
\sech^{g}\left(x+\frac{\Delta}{2}\right)\sech^{g}\left(x-\frac{\Delta}{2}\right) \nonumber\\
& & \times\left[2-\tanh\left(x+\frac{\Delta}{2}\right)+\tanh\left(x
-\frac{\Delta}{2}\right)\right]\!.
\end{eqnarray}
Since the two hyperbolic tangent terms contribute equally, this simplifies to
\begin{equation}
\label{eq-Esech}
E_{\mp} =\mp \frac{g^{2}\Gamma(g+\frac{1}{2})}{\sqrt{\pi}\Gamma(g+1)}\int dx\,
\sech^{g}\left(x+\frac{\Delta}{2}\right)\sech^{g}\left(x-\frac{\Delta}{2}\right)
\left[1+\tanh\left(x-\frac{\Delta}{2}\right)\right]\!.
\end{equation}

We may simplify (\ref{eq-Esech}) using hyperbolic identities, in particular
\begin{equation}
\label{eq-sechid}
\sech(\kappa+\nu)=\frac{\sech\kappa\sech\nu}{1+\tanh\kappa\tanh\nu}.
\end{equation}
Letting $\kappa=x-\frac{\Delta}{2}$ and $\nu=\Delta$ in (\ref{eq-sechid}) gives
\begin{equation}
\label{eq-Ealpha}
E_{\mp}=\mp\frac{g^{2}\Gamma(g+\frac{1}{2})}{\sqrt{\pi}\Gamma(g+1)}\sech^{g}\Delta\int dx\,
\frac{\sech^{2g}\left(x-\frac{\Delta}{2}\right)\left[1+\tanh\left(x-\frac{\Delta}{2}\right)\right]}
{\left[1+\chi\tanh\left(x-\frac{\Delta}{2}\right)\right]^{g}},
\end{equation}
where we have introduced a new dimensionless parameter $\chi\equiv\tanh\Delta$.
Making the substitution $u=\tanh(x-\frac{\Delta}{2})$---the same substitution, we mention,
that makes it possible to calculate the normalization constant in (\ref{eq-normalization})---in
(\ref{eq-Ealpha}) simplifies this further, to
\begin{equation}
\label{eq-EwithC}
E_{\mp}=\mp\frac{g^{2}\Gamma(g+\frac{1}{2})}{\sqrt{\pi}\Gamma(g+1)}\sech^{g}\Delta \int_{-1}^{1}du\,
\frac{(1+u)^{g}(1-u)^{g-1}}{(1+\chi u)^{g}},
\end{equation}
which is a rational function if $g$ is an integer (as, for example, in the supersymmetric case).
Even if $g$ is not an integer, the integral in (\ref{eq-EwithC}) may evaluated analytically in
terms of the hypergeometric function $_{2}F_{1}(a_{1},a_{2};b_{1};z)$. In terms
of $_{2}F_{1}$, the integral is
\begin{equation}
\label{eq-hyperg}
\int_{-1}^{1}du\,\frac{(1+u)^{g}(1-u)^{g-1}}{(1+\chi u)^{g}}=
\sqrt{\pi}\frac{\Gamma(g)}{\Gamma(g+\frac{1}{2})}(1+\chi)^{-g}\,{}
_{2}F_{1}\left(g,g;1+2g;\frac{2\chi}{1+\chi}\right)\!.
\end{equation}
Fortunately, all the $\Gamma$-functions cancel when this is inserted in the expressions for $E_{\mp}$,
leaving
\begin{equation}
\label{eq-Efinal}
E_{\mp}=\mp g\sech^{g}(x_{A}-x_{K1})\left[1+\tanh(x_{A}-x_{K1})\right]^{-g}\,{}
_{2}F_{1}\left(g,g;1+2g;\frac{2\tanh(x_{A}-x_{K1})}{1+\tanh(x_{A}-x_{K1})}\right)\!.
\end{equation}

Since the solitons are far apart,
$\chi\approx1-2e^{-2\Delta}\equiv1-\epsilon$ is close to one,
and it is natural to expand (\ref{eq-hyperg}) around $\chi=1$. If we expand
the integral in (\ref{eq-hyperg}) to ${\cal O}(\epsilon)$, we find,
\begin{equation}
\label{eq-epsilon}
\int_{-1}^{1}du\,\frac{(1+u)^{g}(1-u)^{g-1}}{(1+\chi u)^{g}}=
\int_{-1}^{1}du\,(1-u)^{g-1}+\epsilon\left[g\int_{-1}^{1}du\,\frac{u(1-u)^{g-1}}{(1+u)}\right]\!.
\end{equation}
The ${\cal O}(\epsilon^{0})$ term in (\ref{eq-epsilon}) has value $\frac{2^{g}}{g}$.
However, the ${\cal O}(\epsilon)$ term is divergent,
because $_{2}F_{1}(g,g;1+2g;z)$ is not analytic at $z=1$. Instead, there is additional
logarithmic behavior that needs to be isolated. To find the ${\cal O}(\epsilon\log\epsilon)$
part of the integral, we subtract off the first term on the right-hand side of (\ref{eq-epsilon})
to yield
\begin{equation}
\mathcal{I}=\int_{-1}^{1} du\left[\left(\frac{1+u}{1+\chi u}\right)^{g}-1\right](1-u)^{g-1}.
\end{equation}
The divergent behavior when $\epsilon=0$ comes from the vicinity of the lower limit of integration,
$u=-1$. Substituting $\upsilon=1+\chi u=1+(1-\epsilon)u$ moves the logarithmic behavior to the
dependence on the lower limit,
\begin{equation}
\mathcal{I}=\int_{\epsilon}^{2-\epsilon}\frac{d\upsilon}{1-\epsilon}
\frac{[(\upsilon-\epsilon)/(1-\epsilon)]^{g}
-\upsilon^{g}}{\upsilon^{g}}\left(\frac{2-\epsilon-\upsilon}{1-\epsilon}\right)^{g-1}.
\end{equation}
Since the integrand vanishes at $\epsilon=0$, we may neglect any overall multiplicative
power of $\chi=1-\epsilon$. Moreover, the logarithmic behavior still comes entirely from
the lower limit around $\upsilon=\epsilon$, around which $(2-\epsilon-\upsilon)\approx2$. This
fact also makes the precise value of upper integration limit unimportant; to this order,
we only need the ${\cal O}(\epsilon\log\epsilon)$ part of
\begin{eqnarray}
\mathcal{I} & \sim & \int_{\epsilon}d\upsilon\left[\left(1-\frac{\epsilon}{\upsilon}\right)^{g}-
(1-\epsilon)^{g}\right]2^{g-1} \\
& \approx & 2^{g-1}\int_{\epsilon}d\upsilon\left[\left(1-g\frac{\epsilon}{\upsilon}\right)-
(1-g\epsilon)\right] \\
& \sim & 2^{g-1}\left(g\epsilon\log\epsilon\right).
\end{eqnarray}
The full integral integral for the limit of large separation is therefore
\begin{equation}
\label{eq-epslog}
\int_{-1}^{1}du\,\frac{(1+u)^{g}(1-u)^{g-1}}{(1+\chi u)^{g}}=
\frac{2^{g}}{g}+g2^{g-1}(\epsilon\log\epsilon)+{\cal O}(\epsilon),
\end{equation}
and using (\ref{eq-epslog}) and $\epsilon=2e^{-2\Delta}$, the expression for $E_{\mp}$ out to
${\cal O}(\epsilon\log\epsilon)$ is
\begin{equation}
E_{\mp}\approx\mp g\left[\frac{2^{g}}{\sqrt{\pi}}\frac{\Gamma(g+\frac{1}{2})}
{\Gamma(g+1)}\right]\sech^{g}\Delta\left[1-2\left(g^{2}\Delta\right)e^{-2\Delta}\right]\!,
\label{eq-E-mp}
\end{equation}
which agrees with the earlier results through terms of ${\cal O}(g)$ as $g\rightarrow0$.

Besides the nonanalytic dependence on the small parameter $e^{-(x_{A}-x_{K1})}$, the
prefactor in (\ref{eq-E-mp}) is also nontrivial. The possibility of such a complicated dependence
on $g$ was not countenanced in the original paper~\cite{ref-jackiw} on fermion fractionalization via kink
interactions. The energies, including the prefactor, were correctly stated in Ref.~\cite{ref-goldhaber}
(although only for the
particular supersymmetric value of $g$) and,
subsequent to Ref.~\cite{ref-altschul42}, slightly incorrectly stated
in Ref.~\cite{ref-choi1} (which used an incorrect approximation for the hyperbolic secant).

\section{Stationary System of Kink-Antikink-Kink}

\label{app-3stationary}

Herein, we
shall calculate the required integrals that are needed to get the properties of the almost-zero-energy
fermion modes in a motionless kink-antikink-kink background. The first integral is 
\begin{equation}
    \int dx\,\psi^{*}_{K1}\psi_{K2}=\frac{g}{4^{g}}\int dx\,
    \sech^{g}(x-x_{K1})\sech^{g}(x-x_{K2}).
\end{equation}    
We split up the integration into three regions: 
$\int_{-\infty}^{+\infty}dx\,=\int_{-\infty}^{x_{K1}}dx+\int_{x_{K1}}^{x_{K2}}dx+\int_{x_{K2}}^{+\infty}dx$,
then use (\ref{eq-coshapprox}). This gives
\begin{eqnarray}
    \int dx\,\psi^{*}_{K1}\psi_{K2}
    &=& g\left[\int_{-\infty}^{x_{K1}}dx\,e^{g(2x-x_{K1}-x_{K2})}
    +\int_{x_{K1}}^{x_{K2}}dx\,e^{-g(x_{K2}-x_{K1})}\right. \nonumber\\
    & & \left.+\int_{x_{K2}}^{+\infty}dx\, e^{-g(2x-x_{K1}-x_{K2})}\right] \\
    &=& g\left(\frac{1}{2g}e^{-g\Delta_{2}}+\Delta_{2}e^{-g\Delta_{2}}+\frac{1}{2g}e^{-g\Delta_{2}}\right) \\
    &=& (1+g\Delta_{2})e^{-g\Delta_{2}},
\end{eqnarray}
with the dominant contribution coming from the integration over the region between the kinks, where both
wave functions may be of appreciable magnitude.

We may also verify explicitly that the variational states we identified are indeed stationary (not
mixing under time evolution through the order to which we are working). This means evaluating, for
instance,
\begin{eqnarray}
   \int dx\, \Psi^{\dagger}_{g}H \Psi_{0}&=& \frac{g}{\sqrt{2}}\int dx\left[
   \begin{array}{c}
   \!\psi_{K1}^{*}+b\psi_{K2}^{*}\! \\
   \psi_{A}^{*}
   \end{array}
   \right]^{\!\!T}\!\!\left[
   \begin{array}{c}
   0 \\ \!c(\phi_{A}+\phi_{K2})\psi_{K1}+(\phi_{A}+\phi_{K1})\psi_{K2}\!
   \end{array}\right] \\
   &=& \frac{g}{\sqrt{2}}\int dx\,\psi_{A}\left[c(\phi_{A}+\phi_{K2})\psi_{K1}
   +(\phi_{A}+\phi_{K1})\psi_{K2}\right] \\
   & \equiv & \frac{g}{\sqrt{2}}\left(I_{1}+I_{2}\right).
\end{eqnarray}
We can compute the individual integrals denoted $I_{1}$ and $I_{2}$, keeping that in mind that, as above,
the integration will be dominated by the region between the solitons---in this case, $x_{A}<x<x_{K2}$.
The integrals over other parts of the real line will
be doubly exponentially small, and therefore subject
to being neglected. With the scalar field profiles approximated by signum functions and the fermion
wave functions by the corresponding cusped exponentials,
\begin{eqnarray}
    I_{1} &=& c\int dx\,\psi_{A}(\phi_{A}+\phi_{K2})\psi_{K1} \\ 
    & \approx & -2cg\int_{x_{A}}^{x_{K2}}dx\, e^{-g(2x-x_{A}-x_{K1})} \\
    &=& c\left[e^{-g(2\Delta_{2}-\Delta_{1})}-e^{-g\Delta_{1}}\right] \\
    &=& e^{-g(\Delta_{2}-2\Delta_{1})}\left[e^{-g(2\Delta_{2}-\Delta_{1})}-e^{-g\Delta_{1}}\right] \\
    & \approx & -e^{-g(\Delta_{2}-\Delta_{1})}.
\end{eqnarray}
Proceeding along the same lines,
\begin{eqnarray}
    I_{2} & \approx & \int_{x_{K1}}^{x_{A}}dx\,\psi_{A}(\phi_{A}+\phi_{K1})\psi_{K2} \\    
    & \approx &  e^{-g(\Delta_{2}-\Delta_{1})};
\end{eqnarray}
so the sum $I_{1}+I_{2}$ vanishes to the order of our calculations. 
Hence $\int dx\, \Psi^{\dagger}_{g}H \Psi_{0}=0$; a very similar calculation also reveals
that $\int dx\, \Psi^{\dagger}_{e}H \Psi_{0}=0 $. This leaves one remaining integral to calculate
to verify orthogonality,
\begin{eqnarray}
     \int dx\, \Psi^{\dagger}_{g}H \Psi_{e} &=& \frac{g}{2} \int dx\left[
   \begin{array}{c}
   \!\psi_{K1}^{*}+b\psi_{K2}^{*}\! \\
   \psi_{A}^{\dag}
   \end{array}
   \right]^{\!\!T}\!\!\left[
   \begin{array}{c}-(\phi_{K1}+\phi_{K2})\psi_{A} \\
   (\phi_{A}+\phi_{K2})\psi_{K1}+b(\phi_{A}+\phi_{K1})\psi_{K2}
   \end{array}
   \right] \\
   &=&\frac{g}{2}\left\{-\int dx\,\left(\psi_{K1}+b\psi_{K2}\right)(\phi_{K1}+\phi_{K2})
   \psi_{A}\right. \nonumber\\
   & & \left.+\int dx\,\psi_{A}[(\phi_{A}+\phi_{K2})
   \psi_{K1}+b(\phi_{A}+\phi_{K1})\psi_{K2}]\right\} \\
   & \equiv & \frac{g}{2}\left(I'_{1}+I'_{2}\right).
\end{eqnarray}
Since $b$ is already a small parameter, and $\phi_{K1}(x)+\phi_{K2}(x)=0$ for $x_{K1}<x<x_{K2}$
in the signum function approximation, $I_{1}'$ is dominated by
\begin{equation}
\label{eq-I1-prime}
    I'_{1}\approx 2\int_{-\infty}^{x_{K1}}dx\,\psi_{K1}\psi_{A}
    -2\int_{x_{K2}}^{+\infty}dx\,\psi_{K1}\psi_{A};
\end{equation}
moreover, the second integral on the right-hand side of (\ref{eq-I1-prime}), for $x>x_{K2}$, is clearly
also negligible, leaving $I_{1}'\approx e^{-g\Delta_{1}}$. Finally, $I_{2}'\approx -e^{-g\Delta_{1}}$
may be assembled entirely out of the integrals we have already previously computed;
hence $\int dx\, \Psi^{\dagger}_{e}H \Psi_{g}=0$, as expected.

Again using the integrals we have already evaluated, we can directly compute the energies of the
discrete states, obtaining 
$\int dx\, \Psi^{\dagger}_{e}H \Psi_{e}= g e^{-g\Delta_{1}}$,
$\int dx\, \Psi^{\dagger}_{g}H \Psi_{g}= -g e^{-g\Delta_{1}}$, and (trivially)
$\int dx\, \Psi^{\dagger}_{0}H \Psi_{0}= 0$. If we wish to check that
$\int dx\, \Psi^{\dagger}_{0}H \Psi_{g}= \int dx\, \Psi^{\dagger}_{0}H \Psi_{e}= 0,$
both integrals reduce to
\begin{equation}
    \int dx\, \left(c\psi_{K1}^{*}+\psi_{K2}^{*}\right)
    (\phi_{K1}+\phi_{K2})\psi_{A}= -e^{-g(\Delta_{2}+\Delta_{1})},
\end{equation}
which is negligible to the order we are keeping.

\section{Integrals with a Moving Third Kink}

\label{app-3moving}

Here we compute the integrals relevant to the scenario discussed in section~\ref{sec-3moving},
in which there is a third kink at $x_{K2'}$, moving towards a distant kink-antikink system. The
ground and excited fermion states remain orthogonal, since
\begin{eqnarray}
     \int dx\, \Psi'^{\dagger}_{g}\Psi'_{e} & = &
     -\frac{b'v}{2\sqrt{2}}\int dx\,\Psi_{g}^{\dagger}\sigma_{2}\Psi_{K2'}
     -\frac{b'v}{2\sqrt{2}}\int dx\,\Psi^{\dagger}_{K2'}\sigma_{2}\Psi_{e} \\
     &=& -\frac{b'v}{2\sqrt{2}}\left(\frac{i}{\sqrt{2}}\int dx\,\psi_{A}^{*}\psi_{K2'}+
     \int dx\,\psi_{K2'}^{*}\psi_{A}\right)\nonumber\\
     &=& -\frac{b'v}{2}g\left(\Delta'_{2}-\Delta_{1}\right)e^{-g(\Delta'_{2}-\Delta_{1})} \\
     & \approx & 0,
\end{eqnarray}
because of the exponential smallness of $b'$.
However, the zero-energy state no longer remains instantaneously orthogonal to the other two. Instead,
there is a mixing parameter
\begin{eqnarray}
    -\lambda_{1}\equiv\int dx\, \Psi'^{\dagger}_{g}\Psi'_{0} & = &
    -\frac{v}{2}\int dx\, \Psi_{g}^{\dagger}\sigma_{2}\Psi_{K2'}
    -\frac{bv}{2\sqrt{2}}\int dx\,\Psi^{\dagger}_{K2'}\sigma_{2}\Psi_{0} \\
     &=& -\frac{v}{2}\left[\frac{i}{\sqrt{2}}g\left(\Delta_{2}'-\Delta_{1}\right)
     e^{-g(\Delta_{2}'-\Delta_{1})}\right]-0 \\
     &=& -\frac{igv}{2\sqrt{2}}\left(\Delta_{2}'-\Delta_{1}\right)e^{-g(\Delta_{2}'-\Delta_{1})}.
\end{eqnarray}
We can also compute the overlap with the excited state; unsurprisingly, we find the same
magnitude, $\int dx\,\Psi'^{\dagger}_{e}\Psi'_{0}=\lambda_{1}$.

To determined the time evolution, we also need
to evaluate integrals that have the Hamiltonian sandwiched between the wave functions. These
are needed to simplify the terms in which $H$ appears in (\ref{eq-Dirac-abc}).
To start with, we have
\begin{equation}
    \int dx\,\Psi'^{\dagger}_{e}H \Psi'_{g}=
    -\frac{b'v}{2\sqrt{2}}\int dx\,\Psi_{e}^{\dagger}H\sigma_{2}\Psi_{K2'}
    -\frac{b'v}{2\sqrt{2}}\int dx\,\Psi^{\dagger}_{K2'}\sigma_{2}H\Psi_{g}.
\end{equation}
Computing the second integral on the right-hand side separately, to see what its structure looks like,
we find
\begin{eqnarray}
    \int dx\,\Psi^{\dagger}_{K2'}\sigma_{2}H\Psi_{g} & = &
    \frac{g}{\sqrt{2}} \int dx \left[
   \begin{array}{c}
   0\\
   \!\!-i\psi_{K2'}^{*}\!\!
   \end{array}
   \right]^{\!\!T}\!\!\left[
   \begin{array}{c}
   (\phi_{K1}+\phi_{K2})\psi_{A} \\
   \!\!(\phi_{A}+\phi_{K2})\psi_{K1}+b'(\phi_{A}+\phi_{K1})\psi_{K2}\!\!
   \end{array}
   \right] \\
   &=&-\frac{ig}{\sqrt{2}}\int dx\,\psi_{K2'}[(\phi_{A}+\phi_{K2'})\psi_{K1}+
   b'(\phi_{A}+\phi_{K1})\psi_{K2'}] \\
   &=& -\frac{ig^2}{\sqrt{2}\left(4^{g}\right)}
   \left[-2\int_{x_{A}}^{x_{K2'}}dx\,\sech^{g}(x-x_{K2'})\sech^{g}(x-x_{K1})\right. \nonumber\\
   & & \left.+2b'\int_{x_{K1}}^{x_{A}}dx\,\sech^{2g}(x-x_{K2'})\right] \\
   & \approx & ig^2\sqrt{2}(\Delta_{2}'-\Delta_{1})e^{-g\Delta_{2}'}.
\end{eqnarray}
One can clearly see that $ \int dx\,\Psi'^{\dagger}_{e}H \Psi'_{g}$ is
${\cal O}\left(e^{-2g\Delta_{2}'}\right)$, and hence to our order of calculation,
$\int dx\,\Psi'^{\dagger}_{e}H \Psi'_{g}\approx0$. It is also evident from these integrations that
$\int dx\,\Psi'^{\dagger}_{e}H \Psi'_{e} = \int dx\,\Psi^{\dagger}_{e}H \Psi_{e} = E_{+}= ge^{-g\Delta_{1}}$
and $\int dx\,\Psi'^{\dagger}_{g}H \Psi'_{g}=\int dx\,\Psi^{\dagger}_{g}H \Psi_{g}=-E_{+}=-ge^{-g\Delta_{1}}$,
unchanged up to the order of our calculation.

Similarly, we may evaluate
\begin{eqnarray}
    \int dx\,\Psi'^{\dagger}_{0}H \Psi'_{0} & = &
    -\frac{v}{2}\int dx\,\Psi^{\dagger}_{0}H \sigma_{2}\Psi_{K2'}
    -\frac{v}{2}\int dx\,\Psi^{\dagger}_{K2'}\sigma_{2}H\Psi_{0} \\
    &=& {\cal O}\left(e^{-2g\Delta_{2}}\right)\approx 0.
\end{eqnarray}
Another way to see this result is to note that $\int dx\,\Psi^{\dagger}_{0}H \sigma_{2}\Psi_{K2'}
=(\int dx\,\Psi^{\dagger}_{K2'}\sigma_{2}H\Psi_{0})^{\dagger}$; and since
$\int dx\,\Psi^{\dagger}_{K2'}\sigma_{2}H\Psi_{0}$ is imaginary due to $\sigma_{2}$, their sum will be
zero. However, $\Psi'_{0}$ does have nonzero matrix elements with both excited and ground states---for
example,
\begin{equation}
    \label{eq-Psi0prime-H-Psi0}
    \int dx\,\Psi'^{\dagger}_{0}H \Psi'_{g}=
    -\frac{b'v}{2\sqrt{2}}\int dx\,\Psi_{0}^{\dagger}H\sigma_{2}\Psi_{K2'}
    -\frac{v}{2}\int dx\,\Psi^{\dagger}_{K2'}\sigma_{2}H\Psi_{g}.
\end{equation}
We can conclude that the first term on the right-hand side of (\ref{eq-Psi0prime-H-Psi0})
will not contribute, because the integral
is of order ${\cal O}(e^{-g\Delta_{2}})$, and there is already a factor is $b'$ that makes the whole
expression higher order. On the other hand, the second term is 
\begin{eqnarray}
    \int dx\,\Psi'^{\dagger}_{0}H \Psi'_{g} & \approx &
    -\frac{v}{2}\int dx\,\Psi^{\dagger}_{K2'}\sigma_{2}H\Psi_{g} \\
    &=& -\frac{ig^2v}{\sqrt{2}}(\Delta_{2}'-\Delta_{1})e^{-g\Delta_{2}'},
\end{eqnarray}
and it easy to see that a symmetric relation holds for the excited state,
\begin{equation}
\lambda_{2}\equiv\int dx\, \Psi'^{\dagger}_{0}H\Psi'_{e} = \int dx\,\Psi'^{\dagger}_{0}H \Psi'_{g}=
-\frac{ig^{2}v}{\sqrt{2}}(\Delta_{2}'-\Delta_{1})e^{-g\Delta_{2}'}.
 \end{equation}
 
Finally, we shall calculate the matrix elements of the
time derivative terms that also appear in (\ref{eq-Dirac-abc}),
beginning with
\begin{eqnarray}
    \int dx\,\Psi'^{\dagger}_{e}\partial_{t}\Psi'_{0} & = &
    \frac{1}{\sqrt{2}}\int dx  \left[
    \begin{array}{c}
    \!\psi_{K1}^{*}+b'\psi_{K2'}^{*} \!\\
    -\psi_{A}^{*}
    \end{array}
    \right]^{\!\!T}\!\!\left[
    \begin{array}{c}
    cgv\psi_{K1}+v\partial_{x}\psi_{K2'} \\
    0
    \end{array}
    \right] \\
    &=& \frac{cgv}{\sqrt{2}}+\frac{v}{\sqrt{2}}\int dx\,\psi_{K1}\partial_{x}\psi_{K2'} \\
     &=& \frac{cgv}{\sqrt{2}}+\frac{v}{\sqrt{2}}(d'g) \\
     &=&-\frac{b'gv}{\sqrt{2}}=-\frac{bgv}{\sqrt{2}}+\mathcal{O}(v^{2}).
\end{eqnarray}
Furthermore, we can compute the following term using the integrals we have already done:
\begin{eqnarray}
    \tau & \equiv & \int dx\,\Psi'^{\dagger}_{e}(H-i\partial_{t})\Psi'_{0}
    =\int dx\,\Psi'^{\dagger}_{g}(H-i\partial_{t})\Psi'_{0} \\
    & = & -\int dx\,\Psi'^{\dagger}_{0}(H-i\partial_{t})\Psi'_{e}
    =-\int dx\,\Psi'^{\dagger}_{0}(H-i\partial_{t})\Psi'_{g} \\
    &= & -\lambda_{2}+\frac{ibgv}{\sqrt{2}} \\
    &=& -\frac{igv}{\sqrt{2}}\left(g\Delta_{1}+e^{2g\Delta_{1}}\right)e^{-g\Delta_{2}'}.
    \label{eq-tau-prelim}
\end{eqnarray}
However, at the order we are considering only the second term in parentheses in (\ref{eq-tau-prelim})
actually needs to be retained. We can write
$g\Delta_{1}+e^{2g\Delta_{1}}=e^{2g\Delta_{1}}\left(1+e^{-2g\Delta_{1}}g\Delta_{1}\right)$, which
in this form is clearly equivalent to
\begin{equation}
\label{eq-tau-final}
\tau\approx-\frac{igv}{\sqrt{2}}e^{-g\left(\Delta_{2}'-2\Delta_{1}\right)}.
\end{equation}

Now we also define
\begin{equation}
\lambda\equiv-\tau+E_{+}\lambda_{1}=
\frac{igv}{\sqrt{2}}\left[\frac{g}{2}(\Delta_{1}+\Delta_{2}')+e^{2g\Delta_{1}}\right]e^{-g\Delta_{2}'}.
\end{equation}
However, again to the level of accuracy of we are using,
$-\lambda$ actually takes the same value as $\tau$ (\ref{eq-tau-final}).
Then taking the decomposition (neglecting all but the almost-zero-energy modes),
\begin{equation}
     (H-i\partial_{t})\Psi'_{e}= \alpha_{1}\Psi'_{e}+ \alpha_{2}\Psi'_{0}+\alpha_{3}\Psi'_{g} 
\end{equation}
of $(H-i\partial_{t})\Psi'_{e}$,
we can integrate the equation after multiplying by $\Psi'^{\dagger}_{e}$, $\Psi'^{\dagger}_{0}$ and
$\Psi'^{\dagger}_{g}$ on the left, to give us the three relations,
\begin{eqnarray}
    E_{+}&=& \alpha_{1}+\alpha_{2}\lambda_{1}\\
    -\tau&=& -\alpha_{1}\lambda_{1}+\alpha_{2}+\alpha_{3}\lambda_{1}\\
       0&=& -\alpha_{2}\lambda_{1}+\alpha_{3}.
\end{eqnarray}
Solving these equations and neglecting terms of order ${\cal O}(e^{-2g\Delta_{2}'})$, we find
$\alpha_{3}=0$, $\alpha_{1}=E_{+}$, and thus
$\alpha_{2} =-\tau+E_{+}\lambda_{1}\equiv\lambda=-\tau$. We can perform
similar calculations for the other wave functions, and summarizing the results, we have
\begin{eqnarray}
\label{eq-decom1}
(H-i\partial_{t})\Psi'_{e} & = & E_{+}\Psi'_{e}+ \tau\Psi'_{0} \\
\label{eq-decom2}
(H-i\partial_{t})\Psi'_{g} & = & -E_{+}\Psi'_{g}+ \tau\Psi'_{0} \\
\label{eq-decom3}
(H-i\partial_{t})\Psi'_{0} & = & -\tau\Psi'_{e}- \tau\Psi'_{g}. 
\end{eqnarray}

\section{Solving the Coupled Differential Equations}

\label{app-ODE}

We can write the equations (\ref{eq-final0}--\ref{eq-final0c}) in the matrix form
\begin{equation}
\label{eq-matrix1}
     \partial_{t} \left[
\begin{array}{c}
\alpha \\
\beta \\
\gamma
\end{array}
\right]=-i
\left[
\begin{array}{ccc}
E_{+} & \tau & 0 \\
-\tau & 0 & -\tau \\
0 & \tau & -E_{+}
\end{array}
\right]\!\!
\left[
\begin{array}{c}
\alpha \\
\beta \\
\gamma
\end{array}
\right]\equiv-iA\!
\left[
\begin{array}{c}
\alpha \\
\beta \\
\gamma
\end{array}
\right]\!.
\end{equation}
The matrix appearing in (\ref{eq-matrix1})
can, of course, be exactly diagonalized and thus exponentiated for arbitrary
$E_{+}$ and $\tau$. However, we usually only need the eigenvalues
\begin{eqnarray}
\label{eq-eigenvalue}
\omega_{1}&=&0 \\
\omega_{2}&=&\sqrt{E_{+}^2-2\tau^{2}} \\
\omega_{3}&=&-\sqrt{E_{+}^2-2\tau^{2}}
\end{eqnarray}
to the precisions $\omega_{2}\approx E_{+}$ and $\omega_{3}\approx-E_{+}$, with the 
omitted terms being $\mathcal{O}(e^{-2g\Delta_{2}})$.
Similarly, the corresponding eigenvectors at the requisite level of approximation are
\begin{equation}
u_{1}=\left[
\begin{array}{c}
-\frac{\tau}{E_{+}}\\
1\\
\frac{\tau}{E_{+}}
\end{array}
\right],\quad
u_{2}=\left[
\begin{array}{c}
1 \\
\frac{\omega_{2}-E_{+}}{\tau} \\
\frac{\omega_{2}-E_{+}}{\omega_{2}+E_{+}}
\end{array}
\right]\approx\left[
\begin{array}{c}
1\\
-\frac{\tau}{E_{+}}\\
0
\end{array}
\right],\quad
u_{3}\approx\left[
\begin{array}{c}
0\\
\frac{\tau}{E_{+}}\\
1
\end{array}
\right]\!,
\end{equation}
which are orthogonal [to $\mathcal{O}(v^{2})$] because $\tau$ is purely imaginary.
So we can decompose
\begin{equation}
\left[
\begin{array}{c}
\alpha\\
\beta\\
\gamma
\end{array}
\right]
=\left[\beta+(\alpha-\gamma)\frac{\tau}{E_{+}}\right]u_{1}+\left(\alpha+\beta\frac{\tau}{E_{+}}\right) u_{2}+
\left(\gamma-\beta\frac{\tau}{E_{+}}\right)u_{3}.
\end{equation}
Using this decomposition the system of differential equations (\ref{eq-matrix1}) may be written as 
\begin{eqnarray}
    \left[\dot{\beta}+\left(\dot{\alpha}-\dot{\gamma}\right)\frac{\tau}{E_{+}}\right]\!\!u_{1}
    +\!\left(\!\dot{\alpha}+\dot{\beta}\frac{\tau}{E_{+}}\!\right)\!\!u_{2}
    +\!\left(\!\dot{\gamma}-\dot{\beta}\frac{\tau}{E_{+}}\!\right)\!\!u_{3}
    & = & -iE_{+}\!\left(\!\alpha+\beta\frac{\tau}{E_{+}}\!\right)\!\!u_{2} \nonumber\\
    & & +iE_{+}\!\left(\!\gamma-\beta\frac{\tau}{E_{+}}\!\right)\!\!u_{3}.
    \label{eq-eigenvectorODE}
\end{eqnarray}
The coefficients of each linearly independent eigenvector form a separate differential equation.
If the initial conditions specify that $\beta_{0}$ is much larger than $\alpha_{0}$ and $\gamma_{0}$,
we see from the $u_{1}$ equation that $\dot{\beta}\approx0$, since $u_{1}$ does not appear on the
right-hand side of (\ref{eq-eigenvectorODE}). Thus the occupation amplitude of the zero mode
$\beta(t)\approx\beta_{0}$
remains essentially independent of time. The remaining equations reduce to
\begin{eqnarray}
    \label{eq-alphadot}
    \dot{\alpha} & \approx & -iE_{+}\left(\alpha+\beta_{0}\frac{\tau}{E_{+}}\right) \\
    \label{eq-gammadot}
    \dot{\gamma} & \approx & iE_{+}\left(\gamma-\beta_{0}\frac{\tau}{E_{+}}\right)\!,
\end{eqnarray}
which can be easily integrated to give (\ref{eq-final1}) and (\ref{eq-final2}). Note that although
$\alpha,\gamma\ll\beta$, the two terms in (\ref{eq-alphadot}--\ref{eq-gammadot}) may be comparable in
size.

As alreday noted,
the results may also be obtained from the general solution of the initial value problem with the
exponential of the matrix $A$,
\begin{equation}
\left[
\begin{array}{c}
\alpha(t) \\
\beta(t) \\
\gamma(t)
\end{array}
\right]=e^{-iAt}\left[
\begin{array}{c}
\alpha_{0} \\
\beta_{0} \\
\gamma_{0}
\end{array}
\right]=E(t)\left[E(0)\right]^{-1}\left[
\begin{array}{c}
\alpha_{0} \\
\beta_{0} \\
\gamma_{0}
\end{array}
\right]\!,
\end{equation}
where $E(t)$ is a matrix whose columns are the eigenmode solutions of the original system of
differential equations,
\begin{equation}
E(t)=\left[
\begin{array}{c|c|c}
& & \\
u_{1} & u_{2}e^{-i\omega_{2}t} & u_{3}e^{i\omega_{2}t} \\
& & 
\end{array}
\right]\!.
\end{equation}
Neglecting sub-leading terms then gives the same approximate solutions.

\end{document}